\documentclass[journal]{vgtc}                     

\onlineid{0}

\vgtccategory{Research}

\title{Colour Blinded by the Noise}

\author{%
  \authororcid{Harriet Mason}{0009-0007-4568-8215},
  \authororcid{Rachel Rogers}{0000-0002-4145-9630},
  \authororcid{Alison Kleffner}{0009-0004-1000-5232},
  \authororcid{Dianne Cook}{0000-0002-3813-7155}
}

\authorfooter{
  \item
  	Harriet Mason and Dianne Cook are with the Department of Econometrics and Business Statistics, Monash University.
  \item
  	Rachel Rogers is with the School of Mathematics and Physical Sciences, University of Technology Sydney.
    \item
  	Alison Kleffner is with the Department of Mathematics, Creighton University.
  \item 
  	Harriet Mason is the corresponding author. E-mail: harriet.mason1@monash.edu
}

\abstract{%
  Uncertainty visualisation is important for data transparency,
  especially for map visualisations where data is often aggregated.
  Despite the importance of this area, studies evaluating uncertainty
  visualisation lack consensus and produce conflicting results. This
  work introduces a new evaluation approach for uncertainty
  visualisation that attempts to assess uncertainty as noise, rather
  than signal. We evaluate five methods of visualising uncertainty:
  standard choropleth maps, value/variance bivariate maps,
  value-suppressing uncertainty palettes, overlaid sampling, and
  pixelated sampling maps. Built on principles of implicit testing, we
  put an `uncertainty visualisation' spin on the classic Ishihara
  colourblind test to create a novel test that is able to evaluate
  uncertainty as noise. We compare signal visibility to conventional
  hypothesis tests at various levels of group separation. By building
  our experimental design on top of established graphics theory, we
  isolate the plot components that facilitate successful signal
  suppression and establish foundational theory for the perception of
  uncertainty visualisation.
}


\teaser{
 \centering
 \includegraphics[width=\linewidth, alt={Maps of Australia with different renderings of the states.}]{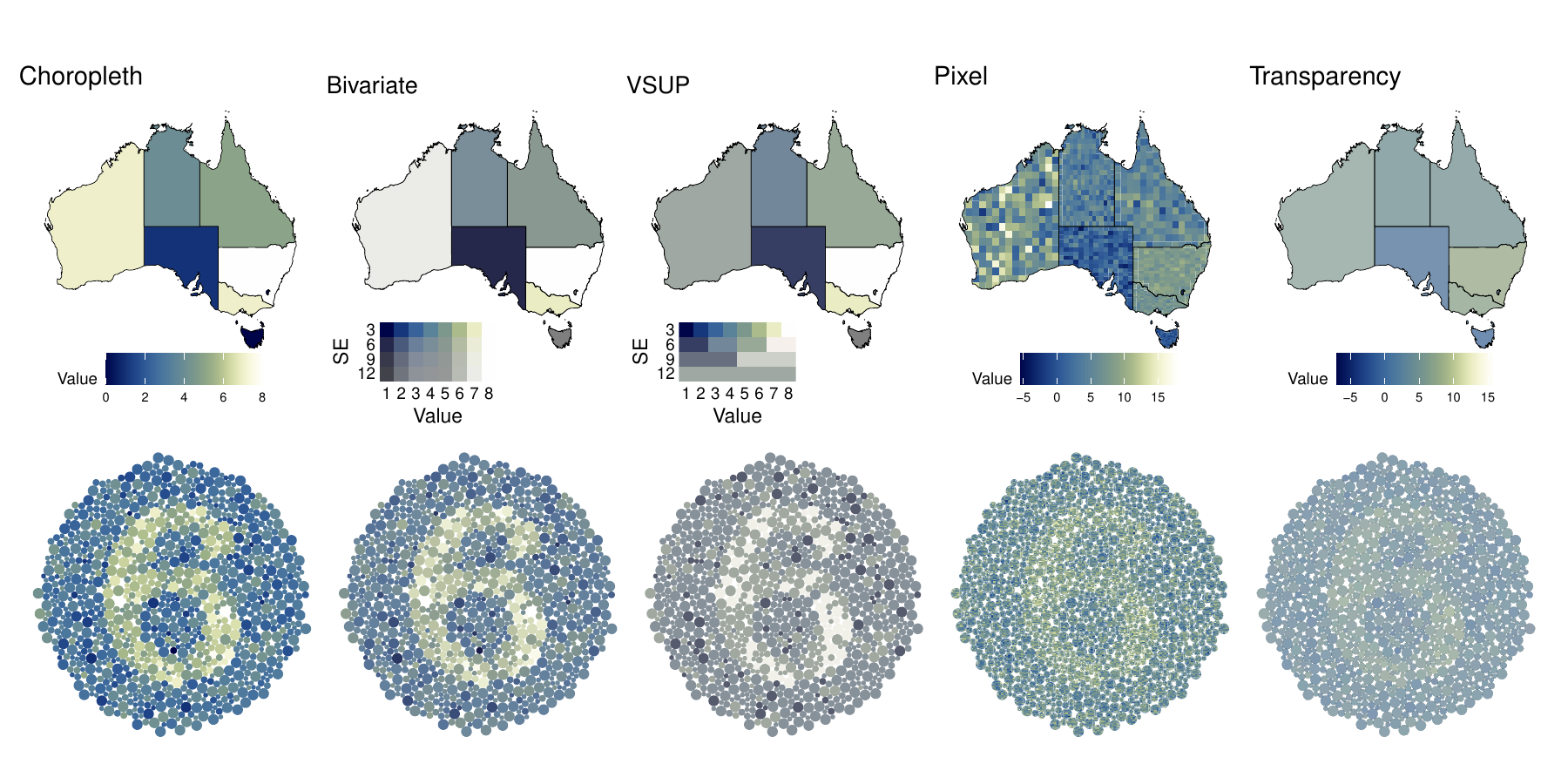}
 \caption{%
 	The five chroropleth maps evaluated in this study, alongside their Ishihara test plate equivalent. Which is the better representation of uncertainty?%
 }
 \label{fig:teaser}
}

\graphicspath{{figs/}{figures/}{pictures/}{images/}{./}} 

\usepackage{tabu}                      
\usepackage{booktabs}                  
\usepackage{lipsum}                    
\usepackage{mwe}                       
\usepackage{ccicons}                   
\usepackage{longtable}
\def\tightlist{}
\usepackage{mathptmx}                  

\usepackage[disable]{backref}

\newcommand{\citeproctext}{}
\newlength{\cslhangindent}
\newlength{\csllabelwidth}
\newlength{\cslentryspacingunit}
\newenvironment{CSLReferences}[2]{}{}
\renewcommand{\bibitem}[2][]{\par\noindent}

\newcommand{\CSLLeftMargin}[1]{\parbox[t]{\csllabelwidth}{#1}}
\newcommand{\CSLRightInline}[1]{\parbox[t]{\linewidth - \csllabelwidth}{#1}\newline}

\renewcommand{\refname}{References}
\renewcommand{\backref}[1]{}
\renewcommand{\backrefalt}[4]{}

\begin{document}

\maketitle


\section{Introduction}\label{introduction}

Uncertainty is routinely present and often ignored in data
visualisation. Because uncertainty can impact the conclusions we draw
about our data, this omission can have negative consequences for our
analysis. This is the key argument behind the ``decision-making''
framework that currently motivates uncertainty visualisation {[}40{]}. A
surprising wealth of motivations nestled under this ``decision making''
umbrella, including: softening unjustified conclusions or discoveries
{[}11{]}, {[}33{]}, {[}60{]}; facilitating rational agent behaviour
{[}61{]}; extracting or comparing probabilities {[}22{]}, {[}26{]}; and
encoding meta information (such as missing values) {[}47{]}. On the
surface, these goals appear harmonious; \emph{however}, an underlying
conflict prevents a visualisation from achieving all of them
simultaneously.

According to Mason et al. {[}36{]}, the current decision-making
framework conflates the two dichotomous roles of uncertainty in
analysis: signal and noise. Signal is the information that we hope to
extract from a plot, while noise is the interference that prevents us
from extracting said signal. These dual roles are a by-product of the
fact that uncertainty visualisations can be described as plots that have
replaced their deterministic data input with a random matrix, where
every individual cell is now a distribution {[}25{]}, {[}36{]}.
Uncertainty acts as signal when we include it in our visualisation to
improve the accuracy of estimates, extract probabilities, elicit
rational decisions, or encode specific meta information. In these cases,
we want information about the individual distributions in and of
themselves, that is, we want to extract ``uncertainty'' statistics such
as \(P(X<x)\) or \(\mathrm{Var}(X)\). Uncertainty acts as noise when we
include it in our visualisation to soften false discoveries or
conclusions. In these cases, we are not interested in extracting
information about each individual distribution, but rather we want to
see how the distribution inputs change our conclusions on the aggregate.
The goal is to imbue our plots with desirable statistical properties
that visually convey the statistical significance of any signal we can
extract from the graphic {[}37{]}.

Despite the fact that visualising uncertainty as noise is one of the
earliest motivations for the field {[}33{]}, the vast majority of
evaluation studies evaluate uncertainty as signal, not noise {[}36{]}.
This is a problem because the role of signal and noise in an analysis
are in direct conflict, so much so that the results of an evaluation
study can be completely flipped on their head if we are not careful with
the distinction {[}36{]}. This accumulates in a literature that is
difficult to synthesise {[}34{]} and studies with antithetical
conclusions, such as the finding that trust is improved when we ignore
uncertainty in high variance scenarios {[}63{]}. Ultimately, this causes
the field to discourage the exact design choices that effectively
visualise uncertainty as noise {[}36{]}. This all leads to the
conclusions that we are in need of a methodology that evaluates
uncertainty as noise, not signal.

The conflicting goals of uncertainty are not the only reason the field
is difficult to synthesise. Kinkeldey et al. {[}28{]} attributes a lot
of the conflict in the research to an engineering approach to
visualisation which implements a usability perspective, rather than
asking why some representations do or do not work. Reducing this
conflict requires a systematic comparison between plots that allow us to
attribute improvements to specific design choices. This becomes far
easier when our scope is narrowed to a specific scenario or plot type.

Maps have been one of the key focal points for uncertainty
visualisation, in part because they offer a particularly challenging
case study, as many familiar statistical visualisation tools become
unavailable when data is referenced on a map {[}57{]}. Cartography also
has a tradition of attention to data quality and a strong desire for
visualisations that ensure the accuracy and reliability of their
conclusions {[}33{]}. Communication of climate events also offers an
intuitive case study in the importance of conveying uncertainty, as we
see evaluation studies ask participants to make decisions about sea
level projections {[}1{]}, flood uncertainties {[}31{]}, wildfire
hazards {[}7{]}, and hurricane forecasts {[}41{]}. While narrowing our
scope down to mapping makes sense given the context of the field,
further simplification is necessary.

A single map may have multiple measurements, with multiple sources of
uncertainty {[}28{]}, {[}34{]}, and trying to test all these sources of
uncertainty at once will overcomplicate our experiment. Effective
evaluations require us to isolate and test marginal units as testing too
many sources of uncertainty at once will be as effective as testing none
at all. Recent research has illustrated that the visual appearance of
uncertainty in a plot is intrinsically linked to the source of
uncertainty in our data {[}37{]}, {[}38{]}, {[}42{]}. Uncertainty in
position looks blurry or sketchy, uncertainty in size looks fuzzy, and
uncertainty in colour looks pixelated {[}32{]} or muddied. For this
reason, isolating the source of uncertainty is equivalent to focusing on
a single aesthetic in a map, such as colour.

Choropleth maps use colour to represent a statistic aggregated over a
region (such as a county, state, or country), with their first uses
dating back over 200 years {[}14{]}. This aggregation is one of the key
sources of uncertainty in maps {[}62{]}. For example, the American
Community Survey is an annual collection of socio-economic data that is
then aggregated over spatial regions; this results in a large margin of
error, which is not always considered in associated visualisations or
analyses {[}23{]}.

This paper describes and implements an experimental approach for
evaluating uncertainty as noise in choropleth maps. Using this
evaluation procedure, we find that historically common methods, such as
bivariate maps, will not prevent the visualisation of statistically
spurious signals while less commonly used resampling methods do.

\section{Background}\label{background}

\subsection{The role of uncertainty as
noise}\label{the-role-of-uncertainty-as-noise}

Visualising uncertainty as noise requires us to make a sort of visual
hypothesis test, where statistically valid signals are visible, while
statistically spurious signals are not. This argument is made explicitly
by MacEachren {[}33{]} and Correll and Gleicher {[}11{]} who argue that
a good uncertainty visualisation should minimise the chance of seeing a
false signal (type I error) or missing a true signal (type II error)
without relying on any statistical expertise from the viewer.

Not only does this goal translate directly from the explicit motivations
of the field, but it is also well established in the underlying
mathematical properties of the plots. If the input of an uncertainty
visualisation is a random matrix {[}25{]}, {[}36{]}, then an uncertainty
visualisation is a function of a random matrix. This means uncertainty
visualisations are visual random variables with convergence properties
that appear as a visual signal materialising from noise as the statistic
converges {[}37{]}. If these plots are random variables, then the
hypothesis test comparison is the natural approach for evaluating their
effectiveness, as hypothesis tests are simply a binary decision made on
a random variable. In the case of uncertainty visualisation, especially
for exploratory data analysis (EDA), the binary outcome is if we ``see''
or ``don't see'' a particular pattern. This property in statistical
graphics, where the statistical strength of a pattern translates to
visibility of signal, is called ``signal modulation'', and more
specifically, ``signal suppression'' when we are primarily concerned
with hiding spurious signals {[}36{]}. For the sake of brevity, we will
use ``uncertainty visualisation'' to exclusively refer to plots made for
signal modulation from this point forward.

\subsection{Lineups and uncertainty
visualisation}\label{lineups-and-uncertainty-visualisation}

Visualisations that seek to align the visibility of a pattern with
classical hypothesis tests are nothing new, and have a well-established
foundation in graphical inference with the lineup protocol {[}5{]},
{[}59{]}. The connection between the two approaches was originally
identified over a decade ago by Hullman et al. {[}22{]}. The lineup
protocol was developed as a visual parallel to hypothesis testing that,
like uncertainty visualisation, ensures the onclusions we draw from our
visualisations are statistically valid. In the lineup protocol, as
described in Buja et al. {[}5{]}, viewers are shown \(M\) plots: \(M-1\)
generated from some null hypothesis, and \(1\) generated from the real
data, called the target plot. Viewers are then asked to identify the
``most different'' plot, where the definition of ``most different'' is
left to the subjective discretion of the viewer. If viewers can
consistently identify the target plot from the lineup, then the null
hypothesis can be rejected, meaning the data used in the target plot is
different from the data used to generate the other \(M-1\) plots.

While both uncertainty visualisations and the lineup protocol provide a
mechanism to perform visual hypothesis testing, their sources of
uncertainty are different. In a lineup, the source of uncertainty is
shown through the null distribution (or future inference), whereas the
uncertainty in an uncertainty visualisation is shown as a feature of the
data itself (typically as distributional inputs {[}25{]}, {[}38{]}). In
classical statistics, the distinction between a distribution around an
estimate and the distribution of a null hypothesis is partially what
differentiates a confidence interval from a hypothesis test. Therefore,
we can think of uncertainty visualisations as the generalised visual
parallel to confidence intervals, while the lineup protocol is the
visual parallel to hypothesis testing. While distinct, both methods of
visual inference can be connected to a statistical hypothesis.

The different sources of uncertainty between lineups and uncertainty
visualisation have a run-on effect in how conclusions are drawn from
each approach. One of these run-on effects is in how the methods show
uncertainty, as lineups show outcomes across multiple plots, while
uncertainty visualisations show the outcomes within a single plot. While
some visualisations disobey this distinction and display the
distribution of a null plot alongside the data {[}16{]}, {[}49{]}, where
the true data is coloured differently from the null hypothesis, this
approach is somewhat antithetical to the goals of both uncertainty
visualisation and the lineup protocol. Uncertainty visualisation and
lineup protocols are built on the idea that statistical significance
should imply visual distinction. If a target plot is significantly
different, it should be visually distinguishable from the null
distribution; if a pattern is statistically significant in an
uncertainty visualisation, it should be visible to viewers. By colouring
the target data differently, the visual differentiability is built into
the plot design and therefore cannot be tested, as would be done in a
lineup scenario.

\subsection{Implicit testing}\label{implicit-testing}

A key benefit of a lineup protocol is its ability to measure a pattern's
visibility without participants explicitly needing to interpret,
understand, or make judgments about the pattern. This is because lineups
leverage implicit testing {[}54{]}. Implicit tests, where participants
must infer the question from the provided stimuli, exist in direct
contrast to explicit tests, which ask users to extract a specific
statistic {[}54{]}. Existing uncertainty visualisation research uses
explicit testing: authors ask participants specific questions or set
goals that the participants must use the visualisation to fulfil. These
studies conflate a participant's ability to see a statistical signal
with outside influences, such as misunderstanding of statistical
concepts, error from cognitive overload, and interference from
participants' prior beliefs or utility functions {[}4{]}, {[}21{]}.
Shifting our evaluation from interpretation to simple signal visibility
with pre-attentive processing will also reduce the cognitive load
required to read the plot {[}54{]}, which is a common issue with
uncertainty visualisations due to their additional complexity {[}4{]}.

While explicit testing is perfectly fine for plots that have been
constructed to showcase a \emph{specific} structure in the data
{[}54{]}, it severely handicaps our ability to evaluate a visualisation
for exploratory data analysis (EDA). Data visualisations, just like
other stimuli, suffer from inattentional blindness, so explicitly
extracting a statistic can disconnect us from the exploratory process
{[}3{]}. This is particularly a problem for uncertainty visualisation in
geography as several authors have expressed a desire for methods that
are suitable for EDA {[}17{]}, {[}34{]}. To evaluate a plot's usefulness
as a tool for EDA, we must evaluate our ability to see a signal before
we even know what that signal might be. Evaluating plots for these
purposes will require a more implicit approach relative to existing
experiments.

Ideally, we would be able to directly translate the lineup protocol to
evaluate different types of uncertainty visualisations. However, because
uncertainty visualisations and lineup protocols are two different
methods of visual inference, this will not be possible. The lineup
protocol can still have a visible pattern in the data when noise itself
generates an interesting pattern (see the LDA lineup in Chowdhury et al
{[}46{]}). This means that a rejection, or failure to reject, in the
lineup protocol does not align with signal visibility, making it an
unsuitable approach for evaluating the specific goals of uncertainty
visualisation.

It is unlikely that we will be able to design a fully implicit test for
uncertainty visualisation, as the assumptions of the approach require
the viewer to bring a null hypothesis with them in the form of an
explicit question. However, we can still use some of the lineup
protocol's design principles as guidance when designing our implicit
tests. One of the unusual design elements of a lineup protocol is that
all the context is removed, as the point is to ``see'' the patterns in
the plot, unhindered by prior beliefs {[}9{]}. Therefore, if we can boil
our question down to such an intuitive level of psychophysical stimuli
that we do not even need scales to interpret it, we can leverage some of
the benefits of the lineup protocol. As a matter of fact, if we do not
need the context of a statistical graphic, we can look outside standard
visualisation evaluation approaches to come up with an effective method.

\subsection{The Ishihara colour blind
test}\label{the-ishihara-colour-blind-test}

At its core, the goal of this uncertainty visualisation experiment is to
measure the effect of noise on signal visibility. To put it another way,
we are trying to measure the effect of a latent variable by measuring
its impact on a primary signal variable. By restricting this to the case
of a choropleth map, we attempt to measure the conditions under which
the latent variable collapses one colour channel (signal) down into
another colour channel (noise). When colour vision deficiency is the
latent variable, this is known as a colour blind test.

The connection between statistical maps and colour blind tests is not
unusual. According to a 1930 review of colour blind tests, methods such
as sorting or matching coloured objects, naming coloured lights, and
distinguishing objects presented in complementary colours have been used
as tests for red-green distinction {[}18{]}, tasks that would not be out
of place in a visual evaluation study. In particular, we are interested
in pseudo-isochromatic colour blind tests, the most popular of which
involve identifying patterns on coloured cards that are visible to those
with standard colour vision, but invisible to those with colour vision
deficiencies {[}18{]}. The Ishihara test is a specific version of a
pseudo-isochromatic test that is commonly used today {[}15{]}, {[}43{]},
where the coloured dots form numbers or paths on a white background
{[}43{]}, {[}51{]}. Notably, these tests use familiar shapes (in the
form of numbers), which provide readily identifiable spatial clusters
without the need for complex definitions. Additionally, it allows us to
perform one of the most under-researched uncertainty visualisation
operations: aggregations of uncertain information over an area {[}28{]}.
The widespread use of this test also lends an air of familiarity to
test-takers, and the simple numerical response reduces the amount of
time needed per trial, allowing for a single individual to view many
plots. Additionally, evidence seems to show that colour blind tests
maintain accuracy when implemented electronically {[}15{]}, {[}27{]}.
Combined, this work suggests that a ``noise'' blind test, which uses
variance rather than colour vision deficiency to make a signal
invisible, would be an effective way to evaluate our choropleth maps.

\subsection{Choropleth maps and the grammar of
graphics}\label{choropleth-maps-and-the-grammar-of-graphics}

Evaluating our graphics using a systematic approach requires a theory of
the ``difference'' in plots. The seminal work of Wilkinson {[}60{]},
\emph{the grammar of graphics}, provides a framework for this, as it
describes the transformation of visualisations as they move through
several layers, which are, in order: data, scales, statistics, geometry,
co-ordinates, and aesthetic mappings. The grammar's influence is seen in
its implementation, as the theory is the foundational structure for
widely used visualisation software like \texttt{ggplot2} {[}58{]} and
Vega Lite {[}48{]}. Using the grammar, visualisation elements can be
incrementally changed, keeping the underlying data constant, and the
effects of these changes can be isolated to particular components of the
grammar {[}54{]}. This approach is usually utilised in conjunction with
the lineup protocol, and the combined pair has been used to compare
polar and cartesian coordinates {[}20{]}, geometric distribution
representations {[}20{]}, and colour palettes {[}44{]}. Until recently,
it would have been difficult (if not impossible) to implement this
approach for uncertainty visualisation, as uncertainty visualisations
were not properly described by the grammar of graphics {[}60{]}.
However, thanks to the recent work {[}25{]}, {[}38{]}, the uncertainty
visualisations used in this study can be fully explained within the
grammar.

There are a few restrictions on the type of choropleth map that will be
evaluated in this experiment. Our first restriction is that we will not
look at maps that primarily change the underlying statistic we are
visualising, such as exceedance probability maps {[}50{]} or Bayesian
surprise maps {[}12{]}. We set this restriction because changing the
underlying statistic changes the fundamental meaning of our plot, rather
than integrating uncertainty into an existing visualisation {[}36{]},
{[}37{]}. Additionally, largely due to computational costs, our approach
is restricted to static plots, ruling out visualisations such as the HOP
{[}22{]}. Finally, to ensure we are comparing variations of the same
plot, we will only compare visualisation that converge to the standard
choropleth map at their deterministic limit {[}37{]}. That is, all the
maps we compare should create an identical choropleth map when the
variance is zero.

\begin{figure}

\centering{

\includegraphics[width=1\linewidth,height=\textheight,keepaspectratio]{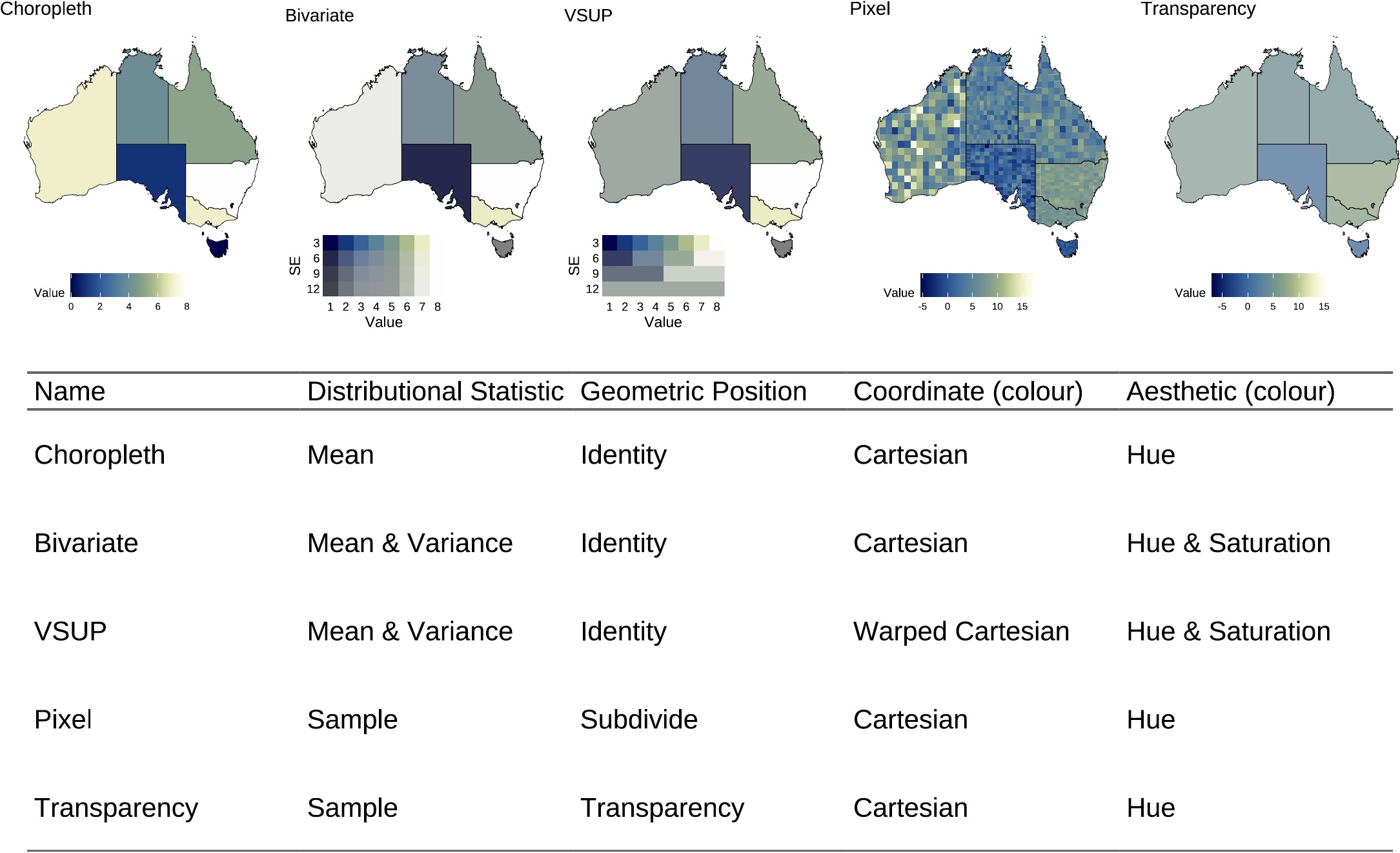}

}

\caption{\label{fig-maps}The five map designs we will be evaluating,
illustrated using the state boundaries of Australia. The distributions
visualised in the maps are the same data, but randomly generated. Along
with an example with each map, we also have the breakdown of how each
map is created, a grammar of graphics breakdown of its important
components. We can use this table to understand how each map diverges
from one another (as well as the standard choropleth map) to understand
how our findings translate to generalisable findings about uncertainty
visualisation.}

\end{figure}

Given these restrictions, this study will consider five map types: the
choropleth, bivariate, Value-Suppressing Uncertainty Palettes (VSUP),
pixel, and transparency maps, shown in Figure~\ref{fig-maps}, alongside
the mapping of their (relevant) components in the grammar of graphics.
All visualisations will have a distribution input for each area, and the
components that are not included in the table are constant for all
visualisations. These map types were chosen not only for their ubiquity,
but also due to what their success or failure can teach us about
uncertainty visualisation. Our goal is to identify the design choices
that lead to effective signal suppression, rather than pick out a
``best'' plot.

The first map in the graphic is the choropleth map, which doesn't
include any uncertainty at all, and will serve as a baseline in the
experiment to compare with other methods. Both Kay {[}25{]} and Mason et
al. {[}38{]} establish that a statistic that represents the distribution
is required for visualising uncertainty, and for cases that ignore
uncertainty, such as the standard choropleth map, we typically represent
a distribution by its mean.

Bivariate colour maps were first used by the US Census Bureau in the
1970s {[}39{]}, and they are one of the older variations of the
choropleth map we will investigate. This approach diverges from the
choropleth map by extracting the variance of the distribution alongside
the mean, and mapping that variance to colour saturation, creating a 2D
colour palette. Mason et al.~Mason at al. {[}36{]} hypothesised that the
bivariate map would be insufficient for signal suppression, as mapping
an estimate and its uncertainty on the parallel channels of hue and
saturation, even if the channels are visually integrable and perceived
as a single unit {[}54{]}, will not introduce enough visual interference
to hide statistically invalid signals.

An alternative approach to the bivariate map is the Value Suppressing
Uncertainty Palette (VSUP) {[}13{]}. Fundamentally, this visualisation
is a bivariate map with a warping coordinate transformation applied to
the colour space {[}60{]}. In the VSUP coordinate space warping, nearby
colours are blended together to reduce their discriminability as the
variance in the estimate increases. Unfortunately, this approach has
significant dependence on the scaling of the variables and the method
taken to blend the colours {[}24{]}. While this method is almost certain
to result in \emph{some} visual interference, there is no reason to
believe that separately extracting a mean and variance, only to
recombine them using a coordinate transformation, would result in
interference that is statistically sound. Uncertainty visualisation is
fundamentally a question about statistical validity, which suggests the
interference should be managed at the statistic level of the grammar.

Rather than change the palette, another alternative to the choropleth
map is the pixel map {[}2{]}, {[}32{]}. This map is identical to the
choropleth map in palette, but represents the distribution as a sample
of \(n\) draws, instead of as a mean and variance. This approach
implements an implicit visualisation of uncertainty {[}10{]} and relies
on human perception to extract estimates such as mean or variance. To
manage overplotting and ensure each draw from the sample is evenly
weighted, the area is subdivided into a grid of outcomes, creating the
pixelated appearance that gives the plot its name. If the overplotting
were managed with animation, we would generate a HOP {[}22{]}. If this
plot is more effective than the VSUP or bivariate, it would suggest that
effective signal suppression requires the visualisation of a full
distribution. This would provide a perceptual argument for the
formalisation of uncertainty as a distribution, as suggested by Kay
{[}25{]}. While we will be using a sample, theoretically, any statistic
that gives a full picture of a distribution (such as a set of quantiles,
the PMF, or the PDF) should have similar results, but investigating
these alternatives is outside the scope of this paper.

Finally, we will test the transparency map, which is identical to the
pixel map, but does not perform the subdivision. Instead, to maintain
equal weighting of every draw, the transparency is set to \(\frac1 n\),
meaning we represent each area using a single colour, but that colour
will not necessarily be able to be matched to a palette. This is not a
problem for the artificial environment of our colour blind tests, but it
would be for standard statistical graphics. As we draw meaning from our
visual signals using the legend of our plot, having values that are
impossible to identify on the scale will leave us unable to interpret
our visual signal. Unlike the other maps, this plot is not a standard
alternative to the choropleth map, but we have included it in our
evaluation as it provides some insight into the design requirements for
uncertainty visualisation. Both the pixel and transparency maps are made
using the \texttt{ggdibbler} {[}38{]} R package, with the only
distinction between them being the position adjustment. Similar plots
can be made in python using \texttt{bootplot} {[}42{]}. The pixel map
diverges from the choropleth map in two ways: through the statistic it
visualises, as well as the number of values it shows. If the pixel map
results in successful signal suppression, the transparency map will
allow us to isolate which of those design changes caused it.

\section{Experimental Design}\label{experimental-design}

\subsection{Hypothesis}\label{hypothesis}

Our experiment has three main hypotheses, listed below.

\textbf{H1}: Mapping standard deviation to a second channel in our
visualisation is insufficient to create signal suppression. If true, the
probability of reading the signal in the bivariate and choropleth maps
will be similar and independent of the standard deviation in our
estimates.

\textbf{H2}: For interference to be statistically valid, it depends on
the visualisation of the correct statistical information, rather than
post-hoc adjustments in later stages of the grammar. If true, the signal
suppression in the VSUP map will not be proportional to statistical
significance of the signal.

\textbf{H3}: The visualisation of a statistic that conveys the full size
of the distribution, even if it is visualised as a single value, is the
primary requirement for signal suppression. If this is true, the
probability of reading the signal in the pixel and transparency maps
will be similar.

\subsection{Stimuli}\label{stimuli}

We generated our pseudo-isochromatic plates using functions from the
\texttt{ishihara} package {[}52{]}. Each ``plate'' of the test was made
up of about 1000 circles with approximately 20\% belonging to the number
group, \(N\), and the remaining 80\% belonging to the standard
background colour, \(B\). The colour of each circle, \(P_i\), is
represented by a continuous value, \(C_i\). The standard assumption in
uncertainty visualisation is that every variable is represented by a
distribution, with its own central value and standard deviation. In this
case, circles are divided into two treatment groups - the number group
and the non-number group. Each treatment has an overall mean (based on
\(D\)) and standard deviation, which is used to calculate the mean value
for each circle (and ensures that circles for each group are not the
exact same color in the choropleth map). In comparison to the Australian
map shown in Figure~\ref{fig-maps}, each circle can be considered as a
state. Each circle is then assigned a distribution based on its
calculated mean and standard deviation (\(V\)) - this is the uncertainty
we wish to visualize.

We replicated this scenario using a Bayesian hierarchical model with
\(C_i\sim~N(\bar{n}_i,V^2)~\forall~P_i~\in~N\), or
\(C_i\sim~N(\bar{b}_i,~V^2)~\forall~P_i~\in B\), where \(\bar{n}_i\) is
an outcome from \(\bar N \sim N(0.5*D, 1)\) and \(\bar{b}_i\) is an
outcome from \(\bar B \sim N(-0.5*D, 1)\). This allows us to replicate
the appearance of a standard Ishihara test, even without variance in the
individual estimates, as the colours are mottled by the distribution of
the means. To keep the experiment simple and reduce the confounding
signal from the variance, we kept the standard deviation constant for
all \(C_i\) within a plate. The relative distance between these two
groups, \(D\), as well as the standard deviation within each
observation, \(V\), were factors of interest when generating the data,
as both affect the visibility of the numbers. This resulted in three
factors, each with five levels:

\begin{itemize}
\tightlist
\item
  The five plot types of interest: choropleth, bivariate, VSUP, pixel,
  and transparency.
\item
  The group difference, \(D\), with values of 0, 1, 2, 3, and 4 were
  chosen, where 0 results in no difference between the two groups.
\item
  The standard deviation within each observation, \(V\), was set at 1,
  2, 3, 4, and 5.
\end{itemize}

This resulted in a 5x5x5 factorial design, for a total of 125
experimental plots. Additionally, as the type of number displayed may
affect the readability of the plot, the number displayed in each plot
was randomly assigned for each participant. The order of plots was
completely randomised for each participant.

\begin{figure}

\begin{minipage}[t]{0.33\linewidth}

\centering{

\includegraphics[width=0.9\linewidth,height=\textheight,keepaspectratio]{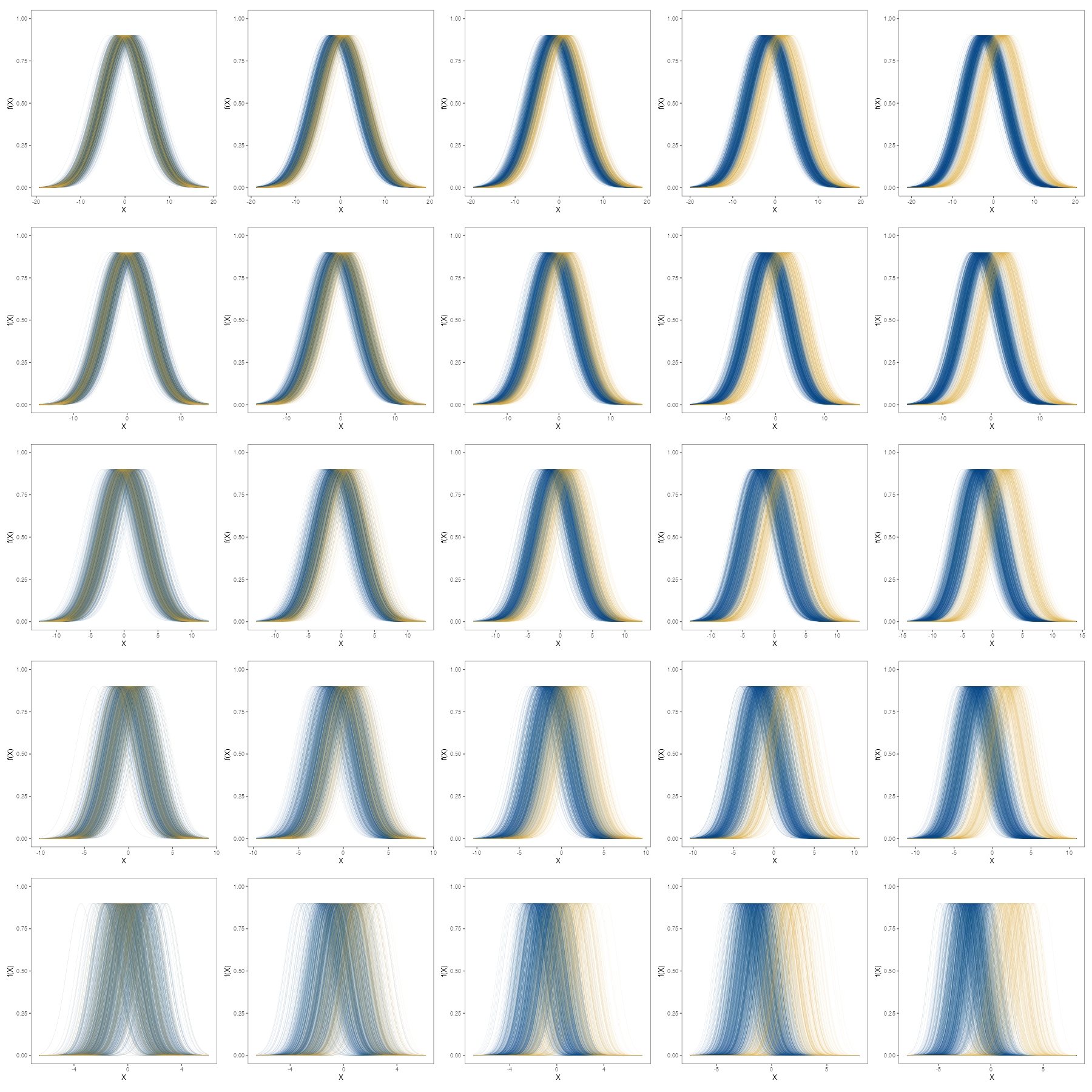}

}

\subcaption{\label{fig-dist25}Theoretical distributions}

\end{minipage}%
\begin{minipage}[t]{0.33\linewidth}

\centering{

\includegraphics[width=0.9\linewidth,height=\textheight,keepaspectratio]{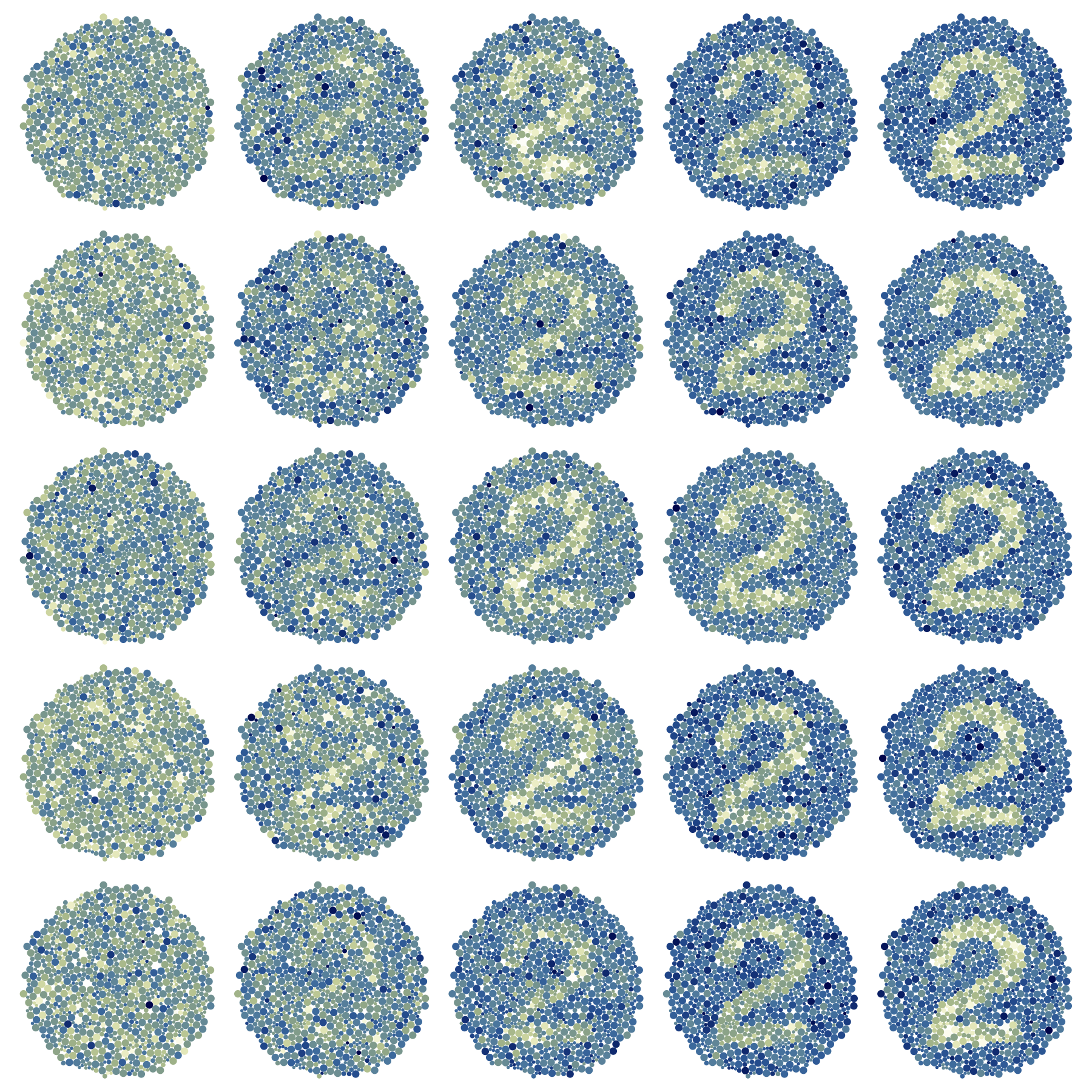}

}

\subcaption{\label{fig-choro25}Choropleth map}

\end{minipage}%
\begin{minipage}[t]{0.33\linewidth}

\centering{

\includegraphics[width=0.9\linewidth,height=\textheight,keepaspectratio]{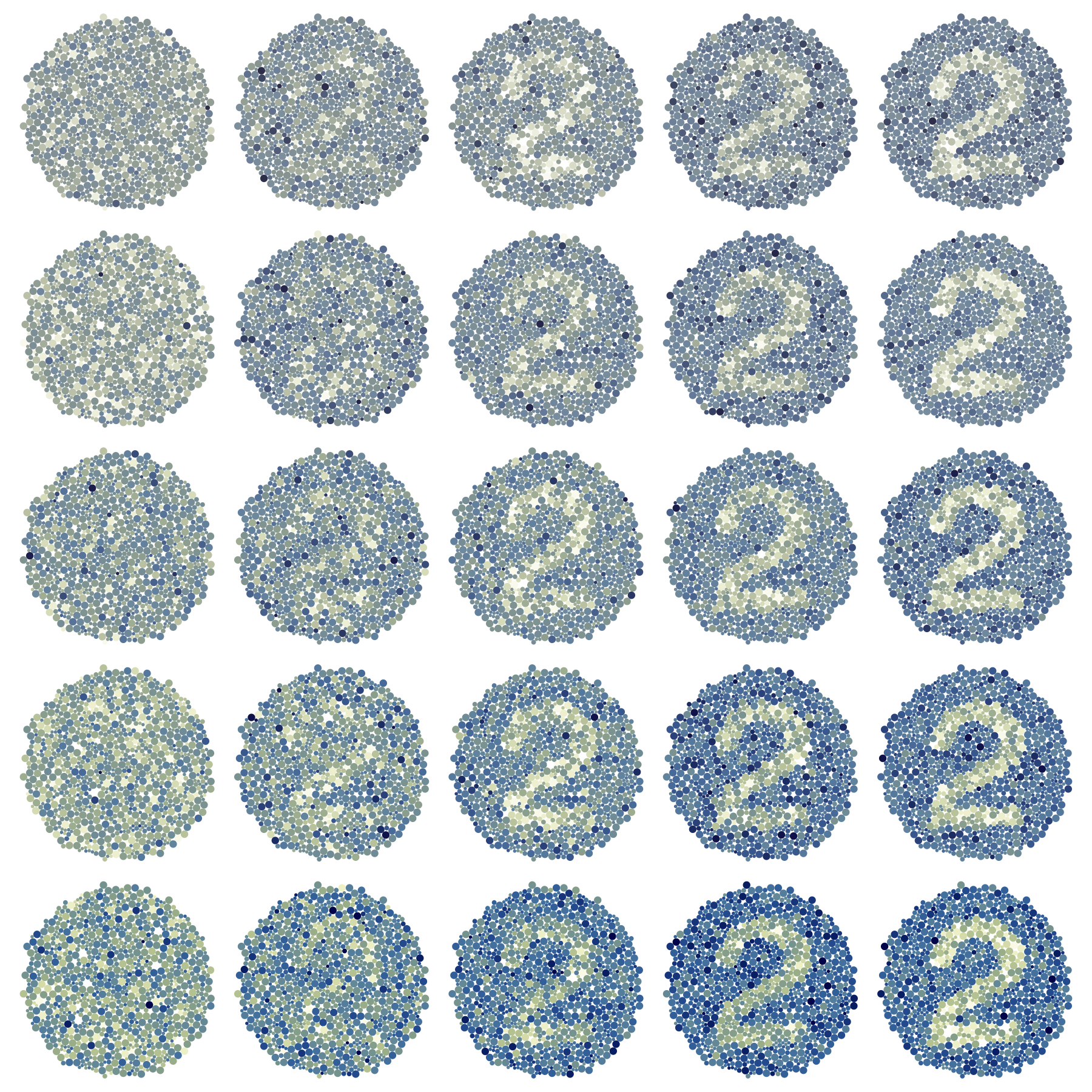}

}

\subcaption{\label{fig-vsup25}Bivariate map}

\end{minipage}%
\newline
\begin{minipage}[t]{0.33\linewidth}

\centering{

\includegraphics[width=0.9\linewidth,height=\textheight,keepaspectratio]{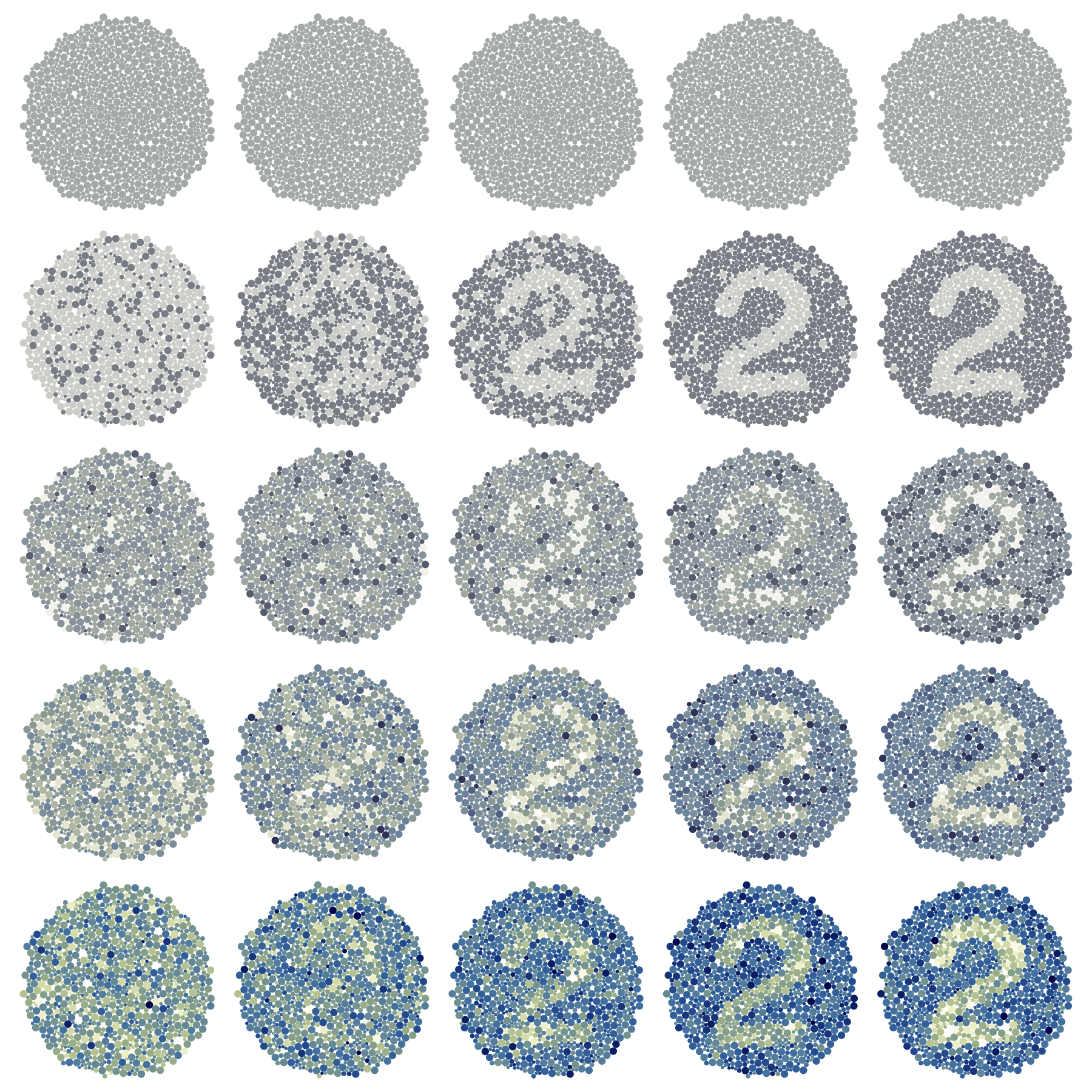}

}

\subcaption{\label{fig-vsup25}VSUP map}

\end{minipage}%
\begin{minipage}[t]{0.33\linewidth}

\centering{

\includegraphics[width=0.9\linewidth,height=\textheight,keepaspectratio]{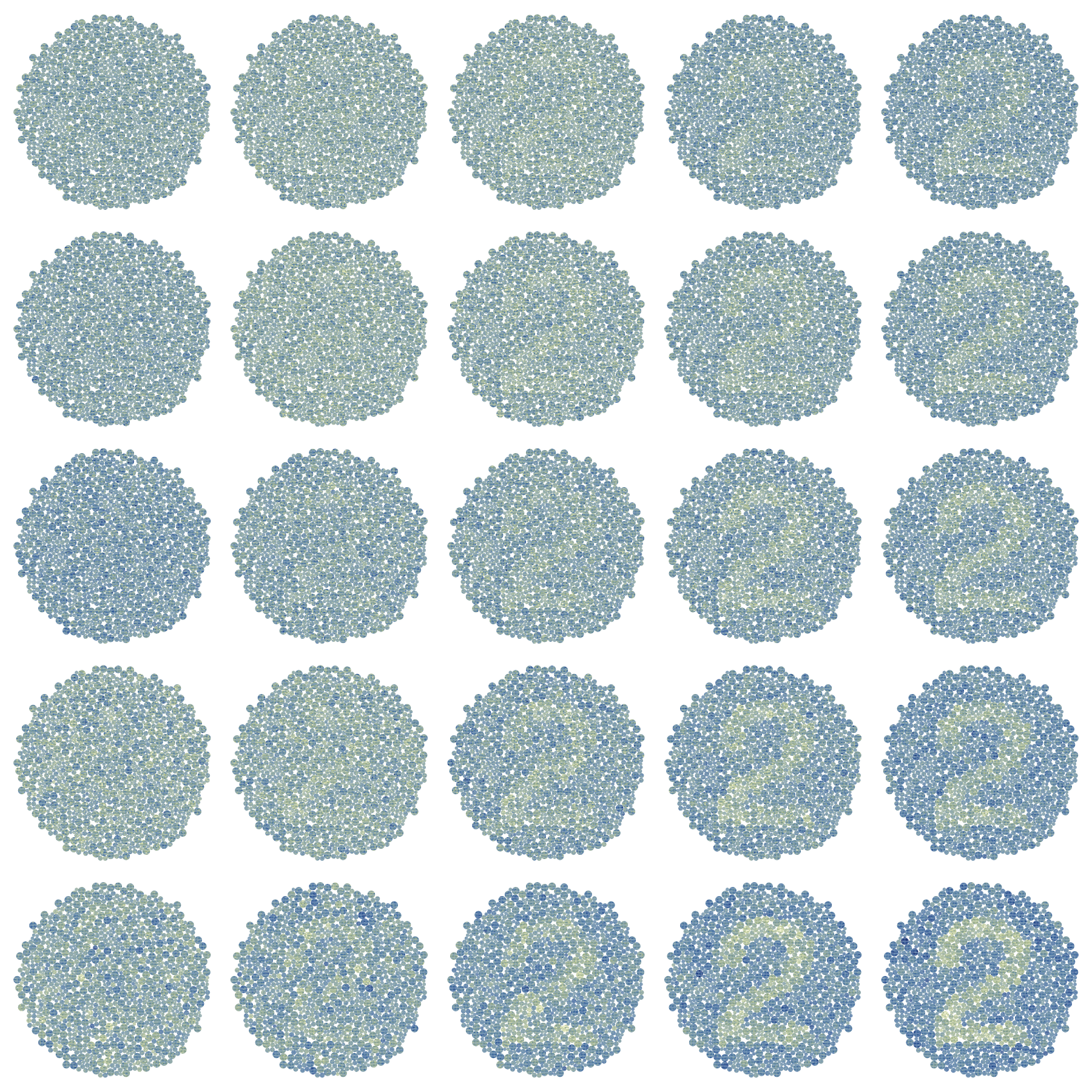}

}

\subcaption{\label{fig-pix25}Pixel map}

\end{minipage}%
\begin{minipage}[t]{0.33\linewidth}

\centering{

\includegraphics[width=0.9\linewidth,height=\textheight,keepaspectratio]{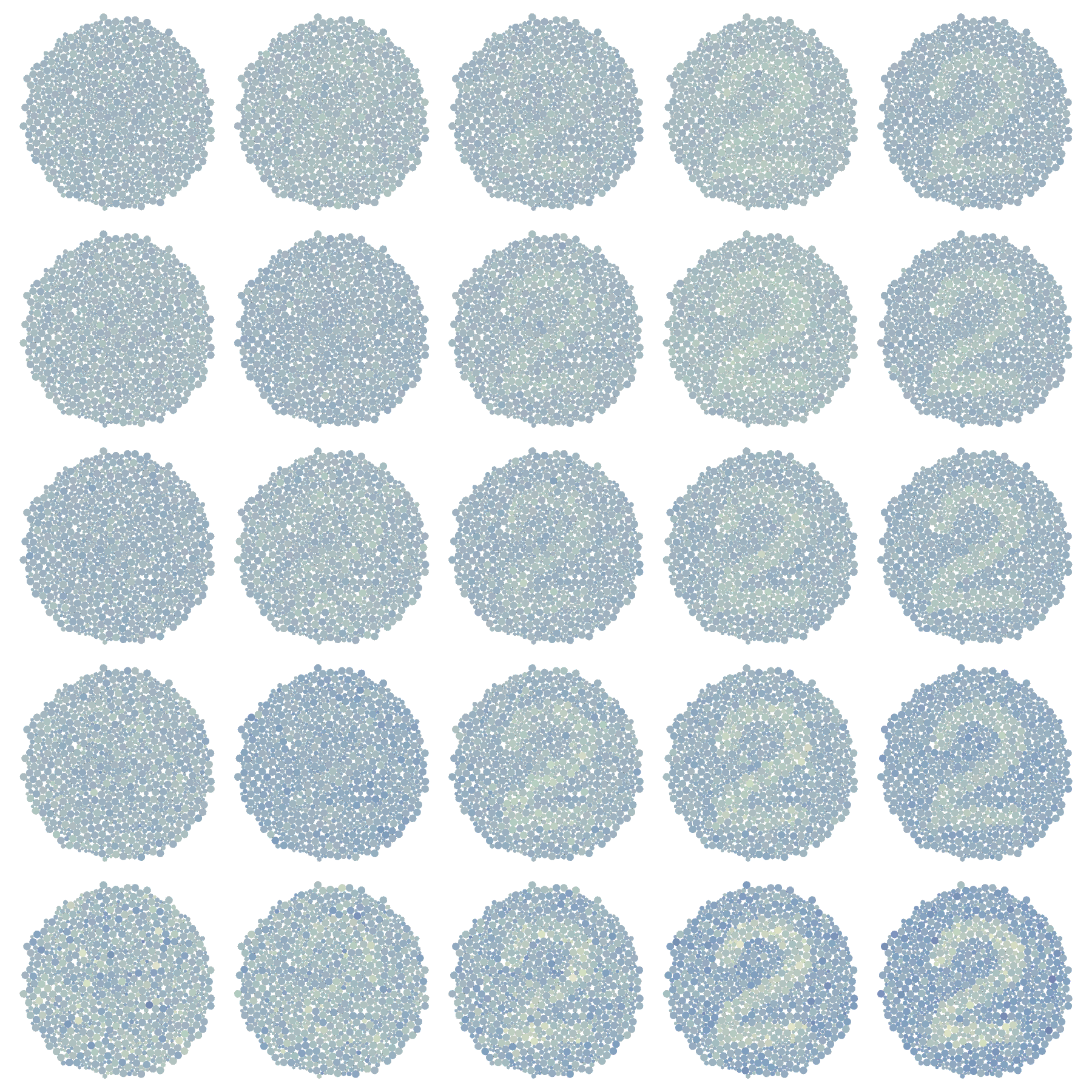}

}

\subcaption{\label{fig-trans25}Transparency map}

\end{minipage}%

\caption{\label{fig-plotexample}Illustration of the experimental design.
Plot (a) shows the data generating distributions, which are effectively
Gaussian with different means and standard distributions for each dot in
the Ishihara plate (yellow indicates distribution for the number, and
blue indicates background dot). Plots (b-f) show the 5x5x5 factorial
design: each one shows plates for the five plot types, where columns
correspond to \(D\) increasing from left to right (0,1,2,3,4), and rows
to \(V\) (1,2,3,4,5) increasing from bottom to top. Plates in the bottom
right corner (\(D=5\) and \(V=1\)) have most visible numbers, and the
top left (\(D=0\), \(V=5\)) is least visible for all plot types.}

\end{figure}%

\subsection{Task and procedure}\label{task-and-procedure}

\begin{figure}

\centering{

\includegraphics[width=1\linewidth,height=\textheight,keepaspectratio]{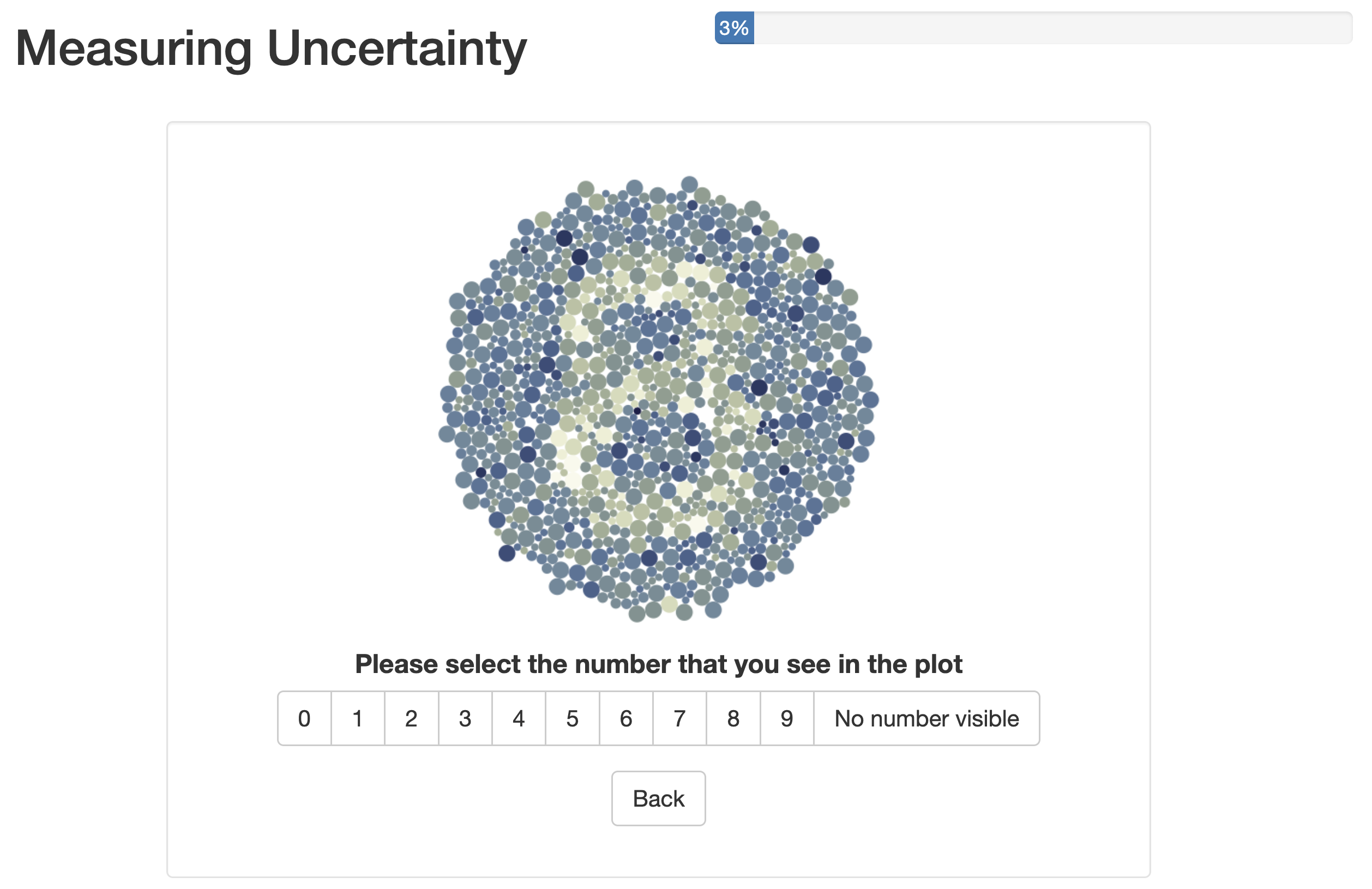}

}

\caption{\label{fig-screenshot}A screenshot of the app as it appeared to
participants while answering the stimulus question.}

\end{figure}

Participants were asked to specify their country of residence, age,
pronouns, education level (based on the International Standard
Classification of Education {[}53{]}), and whether or not they had
colour vision deficiency. Participants were then instructed that they
would be asked to identify numbers in a plot, and that, although the
test may resemble a colour blind test, the test was not meant to measure
colour vision deficiency. They were asked to set their screen brightness
to at least 75\% and to turn off colour filters to ensure consistency in
results {[}15{]}. When viewing the plots, participants were able to
select a digit ranging from 0 to 9, or indicate that no number was
visible within the plot using the interface shown in
Figure~\ref{fig-screenshot}. The study automatically advanced after each
response to speed up survey time, with a back button allowing them to
correct mistaken selections and encourage engagement {[}19{]}. After
every 25 plots, they were shown a plot with a clearly visible number as
an attention check. At the end of the study, participants were able to
leave comments. The study was estimated to take approximately 10 to 15
minutes to complete, and responses were collected via Prolific, an
online survey website.

\begin{figure}

\begin{minipage}[t]{0.10\linewidth}

\centering{

\includegraphics[width=0.85\linewidth,height=\textheight,keepaspectratio]{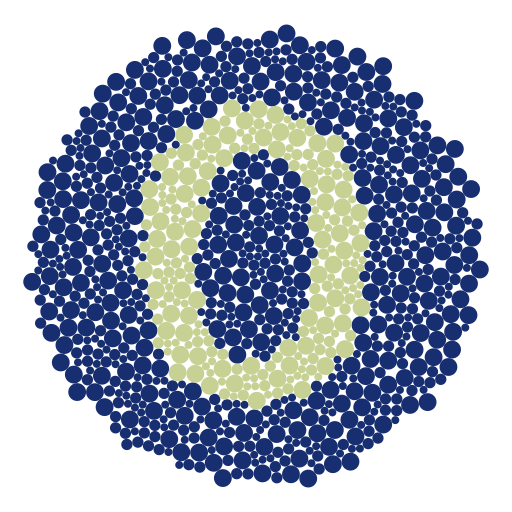}

}

\subcaption{\label{fig-chor}}

\end{minipage}%
\begin{minipage}[t]{0.10\linewidth}

\centering{

\includegraphics[width=0.85\linewidth,height=\textheight,keepaspectratio]{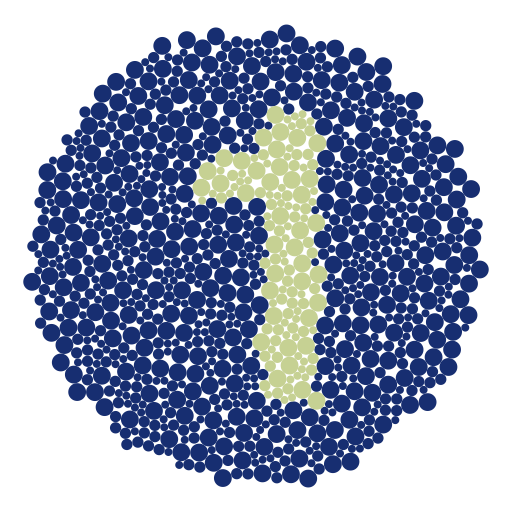}

}

\subcaption{\label{fig-chor}}

\end{minipage}%
\begin{minipage}[t]{0.10\linewidth}

\centering{

\includegraphics[width=0.85\linewidth,height=\textheight,keepaspectratio]{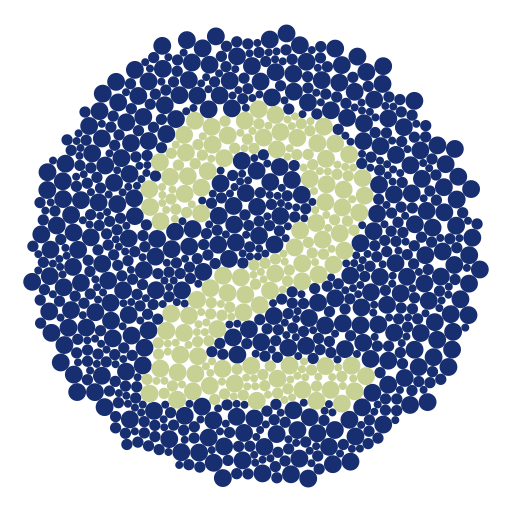}

}

\subcaption{\label{fig-chor}}

\end{minipage}%
\begin{minipage}[t]{0.10\linewidth}

\centering{

\includegraphics[width=0.85\linewidth,height=\textheight,keepaspectratio]{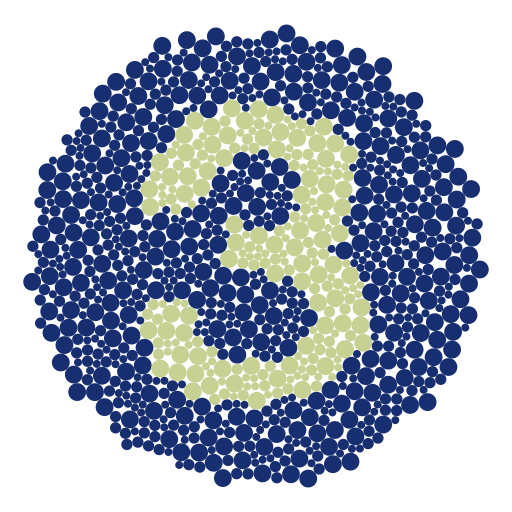}

}

\subcaption{\label{fig-chor}}

\end{minipage}%
\begin{minipage}[t]{0.10\linewidth}

\centering{

\includegraphics[width=0.85\linewidth,height=\textheight,keepaspectratio]{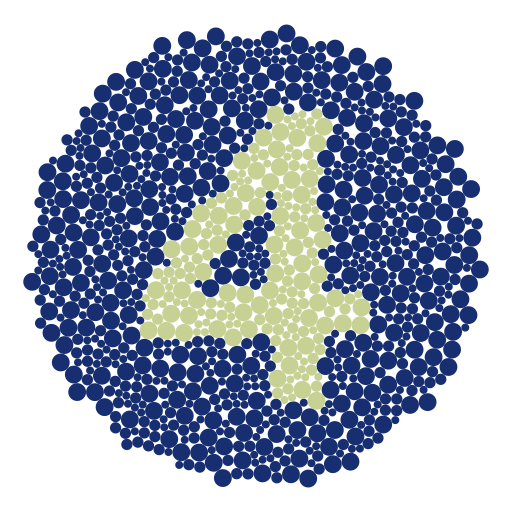}

}

\subcaption{\label{fig-chor}}

\end{minipage}%
\begin{minipage}[t]{0.10\linewidth}

\centering{

\includegraphics[width=0.85\linewidth,height=\textheight,keepaspectratio]{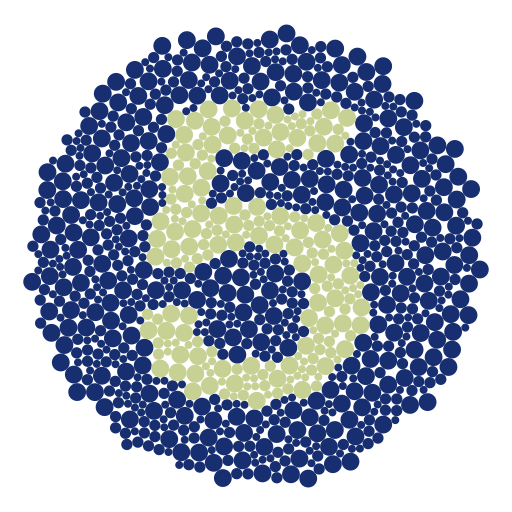}

}

\subcaption{\label{fig-chor}}

\end{minipage}%
\begin{minipage}[t]{0.10\linewidth}

\centering{

\includegraphics[width=0.85\linewidth,height=\textheight,keepaspectratio]{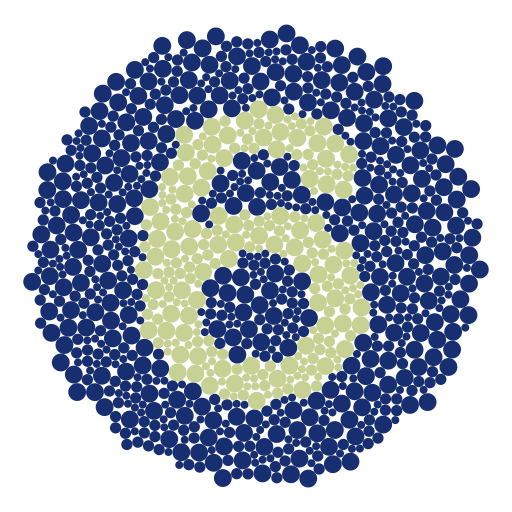}

}

\subcaption{\label{fig-chor}}

\end{minipage}%
\begin{minipage}[t]{0.10\linewidth}

\centering{

\includegraphics[width=0.85\linewidth,height=\textheight,keepaspectratio]{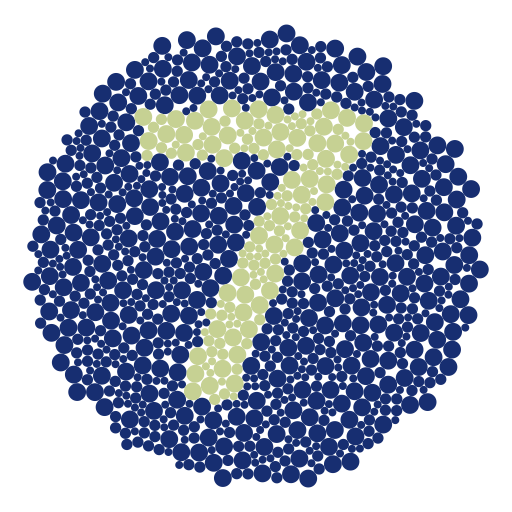}

}

\subcaption{\label{fig-chor}}

\end{minipage}%
\begin{minipage}[t]{0.10\linewidth}

\centering{

\includegraphics[width=0.85\linewidth,height=\textheight,keepaspectratio]{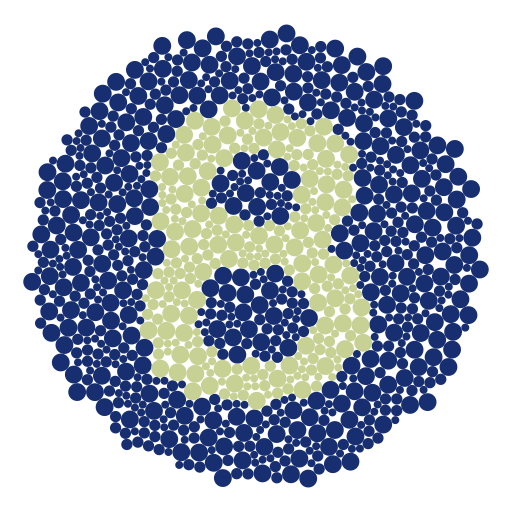}

}

\subcaption{\label{fig-chor}}

\end{minipage}%
\begin{minipage}[t]{0.10\linewidth}

\centering{

\includegraphics[width=0.85\linewidth,height=\textheight,keepaspectratio]{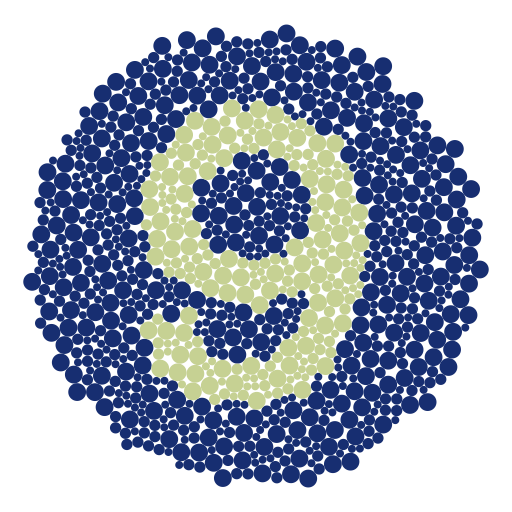}

}

\subcaption{\label{fig-chor}}

\end{minipage}%

\caption{\label{fig-plotexample}The ten attention checks used in the
experiment, included here to illustrate the appearance of all 10 number
plates, as well as the clarity of the attention checks.}

\end{figure}%

General demographic information is meant to provide an overview of our
study sample, while information regarding colour vision deficiency, when
combined with the results, allows us to evaluate unintentional issues in
colour selection and plot design. The number of correct selections will
be compared across plot types and at the selected treatment distances
(\(D\)) and standard deviations (\(V\)). Ultimately, these values will
be compared to generated significance tests in order to compare user
output to accepted measures of difference between groups.

\subsection{Statistical methods}\label{statistical-methods}

\subsubsection{Deciding on a ground
truth}\label{deciding-on-a-ground-truth}

The extensive literature on the lineup means that there is a wealth of
information we can utilise to design our uncertainty visualisation
experiment. Designing an experiment where we have an equivalent
classical test that can be easily calculated is actually the worst-case
scenario for the lineup {[}35{]}, and by extension, uncertainty
visualisation. When we compare our uncertainty visualisations to
classical statistical tests, the idea is that, if the visualisations
behave well when we \emph{do} know the equivalent hypothesis test, they
should also behave well when we don't {[}35{]}. Or even better, they
will work when we don't even know what we want to test, which is the
case for exploratory data analysis.

The correct null hypothesis for any one uncertainty visualisation is not
set in stone, and can depend on the rationale participants use to
identify the numbers. Strictly looking at the data-generating process,
we can see that the data will fulfill the assumptions of a \(t\)-test,
the classic test for a difference between two groups. This would suggest
a \(t\)-test is the appropriate ground truth for our experiment.
However, in the context of a map, participants might use clustering of
lighter or darker spots to make a guess at a shape, even if the full
shape cannot be identified. In these cases, tests of spatial
autocorrelation, such as Moran's I and Geary's C {[}8{]}, that measure
clustering of similar data values in neighboring areal regions (positive
spatial autocorrelation), would be the more appropriate choice. One test
is more appropriate given the data-generating context, but the other is
more appropriate given the spatial context in which the participants
read the plot. For the sake of completeness, we will evaluate
participants' responses against both models.

Using accuracy to compare the plots and hypothesis tests might feel like
a natural approach to comparison, but it would be naive as it would end
up penalising visualisations that do exactly what we have designed them
to do, hide statistically spurious signal. This approach would always
suggest our choropleth map to be the best approach, for the very reason
we want to avoid using it in the first place. We are not looking for
across-the-board ``accuracy'', but rather, a visualisation that is
accurate when a statistical test would be accurate, and inaccurate when
a statistical test would be inaccurate. Therefore, we opt to compare
visualisations using power curves, which is the same approach taken by
the lineup literature {[}35{]}. Power curves show the probability of a
statistical test detecting an effect, given that the effect actually
exists, across a range of effect sizes. Power curves allow us to compare
the efficiency of different tests, so, for a given significance level,
\(\alpha\), a higher power curve (a higher probability of correctly
rejecting a false hypothesis) makes a better test. Ultimately, our goal
is for our test to minimise error (type I or type II), and power curves
allow us to compare hypothesis tests on this metric. Often, hypothesis
tests will cross, so there is no ``uniformly most powerful'' test. An
additional rule of thumb for comparison is ``the steeper the curve, the
better'', as that indicates the test has high sensitivity. Usually, this
power curve analysis is performed using effect size, but effect size can
be difficult to calculate for many statistical tests, and some tests,
such as Moran's I, have no effect size equivalent at all. Therefore, we
used \(D\) and \(V\) as proxies for effect size; however, comparing
signal suppression methods using effect size might create smoother
results, and is a potential future area of research.

\subsubsection{Experimental power curve}\label{experimental-power-curve}

Participant ability to select the correct number in the plot was
modelled using a generalised linear mixed model with a binomial
response. If we let \(Y\) represent the event where a participant
correctly identifies the number in the plot, then the power curve
estimates \(P(Y)\). Participant correctness is treated as the response
variable, and \(D\), \(V\), and plot type are treated as explanatory
variables, with up to the three-way interaction between these variables
included in the model. \(D\) and \(V\) are both continuous, while the
plot type is discrete. It is standard to account for individual ability
to read plots using a random block effect {[}35{]}, so our model also
includes participants as a random effect. The number in the plate is
likely to have a similar impact, so it is also included as a random
effect. The random effects model can be written as follows: \[
logit P(Y_{ijklm})=\eta+\delta_i+\nu_j+\tau_k+\delta_i\nu_j+\delta_i\tau_k+\nu_j\tau_k+\delta_i\nu_j\tau_k+n_l+p_m
\] Here, \(\eta\) is the intercept, \(\delta_i\) is the fixed effect of
distance, \(\nu_j\) is the fixed effect of standard deviation,
\(\tau_k\) is the fixed effect of plot type,
\(n_l\sim N(0,\sigma^2_{number})\) is the random effect for the number
displayed, and \(p_m\sim N(0,\sigma^2_{participant})\) is the random
effect for participant.

\subsubsection{Theoretical power curves}\label{theoretical-power-curves}

Hypothesis tests are usually specified in terms of a test statistic,
which is a function of a sample {[}6{]}. This is a bit of a problem,
because uncertainty visualisations, by their very nature, are designed
for situations where we \emph{don't} have a a precisely observed sample.
Instead, each value is a distribution. There are \(t\)-test equivalents,
such as the pooled \(t\)-test, that allow us to compare two
distributions, but this does not exist for our spatial tests, and
certainly not with the Bayesian hierarchical model we have used to
generate the data. That is, common spatial autocorrelation metrics fail
to take uncertainty into account in their calculations by assuming the
variance of the areal estimates is equal, resulting in biased estimates
of the spatial structure {[}23{]}, {[}29{]}, {[}56{]}. Few alternatives
have been suggested to incorporate uncertainty within spatial
autocorrelation metrics {[}29{]}, {[}56{]}. However, integrating
uncertainty using these theoretical approaches will often only work for
one test or the other and cannot be implemented with both the \(t\)-test
and Moran's I test simultaneously.

Because our data-generating process was known, we used it to perform a
Monte Carlo simulation, where we generated 100 samples of each of the
250 data sets used in the experiment. For each simulated data set,
Moran's I and its corresponding \(p\)-value were calculated for the
alternative hypothesis of positive spatial autocorrelation (spatial
clustering). A permutation test was used for the Moran's I calculation
to establish the null distribution of the I statistic. The \(t\)-test
and its corresponding \(p\)-value were also calculated, with an
alternative hypothesis of a difference between the two means.
Theoretical power curves are calculated from these \(p\)-values.

Estimating power curves for the theoretical hypothesis tests is not
straightforward, so we will adapt the methods used by Li et al.
{[}30{]}. When comparing hypothesis tests, it is standard to restrict
the comparison to tests that have the same Type I error probability
{[}6{]}. Using the null plots, that is, when \(D=0\), we can estimate
the probability of rejecting \(H_0\) when \(H_0\) is true, which we
denote as \(\hat\alpha_k\), for each plot type, \(k\). Using this
\(\hat\alpha_k\) value, we can set the \(\alpha\) for our theoretical
tests, turning our \(p\)-values into accept/reject outcomes. These data
points are comparable to the accept/reject outcomes we were able to
observe in the participants' responses. We then use these data points to
model the theoretical power of the test using a generalised linear mixed
model of the response variable, where \(D\) and \(V\) are treated as
explanatory variables, with a two-way interaction, and \(\hat\alpha_k\)
is specified as the value when \(D=0\). The correct number is also
included as a random effect. This results in the following model: \[
logit P(Y_{ijklm})=\eta+\delta_i+\nu_j+\delta_i\nu_j+n_l
\] Here, \(\eta\) is the intercept, \(\delta_i\) is the fixed effect of
distance, \(\nu_j\) is the fixed effect of standard deviation, and
\(n_l\sim N(0,\sigma^2_{number})\) is the random effect associated with
the number displayed.

\section{Results}\label{results}

\subsection{Participant Information}\label{participant-information}

137 individuals completed the study and passed the attention check. The
median study duration was 8.18 minutes with an interquartile range of
4.72 minutes. Because individuals were provided a `back' button in the
case of accidental submissions, only the last observation entered by an
individual for each plot is considered. A Pearson's chi-squared test for
independence indicates a non-significant relationship between plot type
and number present in the data, indicating that random assignment of
numbers was successful (\(p\)-value of 0.92).

Demographically, participants tended to be younger, use he/him pronouns,
and have a tertiary education. 37.96\% of individuals identified as
being between 18 and 25, while 38.69\% identified as 26 to 35; the
remaining individuals identified as older than 35. 64.96\% of
individuals use he/him pronouns, 32.85\% use she/her, and the remaining
individuals use they/them, or prefer not to answer. 86.86\% of
individuals indicated that their highest education level was tertiary,
while the remaining individuals identified their education level as
below tertiary or post-secondary non-tertiary education. A large number
of participants were from Africa and Europe (41.61\% and 30.66\%,
respectively), and 5.11\% of participants indicated that they were
colour blind or unsure.

\subsection{Results Overview}\label{results-overview}

\begin{figure}

\centering{

\includegraphics[width=0.9\linewidth,height=\textheight,keepaspectratio]{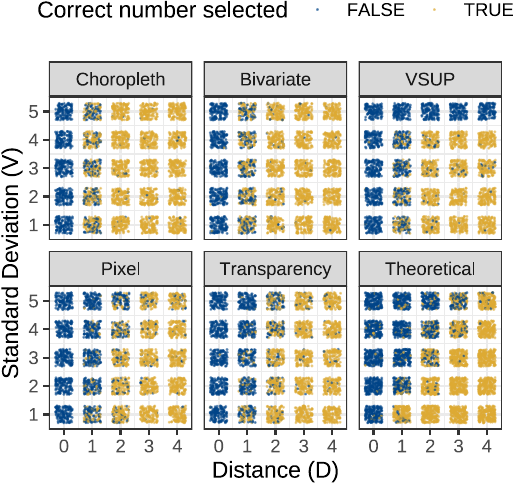}

}

\caption{\label{fig-tileplot}The full set of participant responses,
coloured by their ability to identify the correct number in the plot,
grouped by the distance (\(D\)) between the distributions, and the
standard deviation (\(V\)) in each individual estimate. The theoretical
simulation from the \(t\)-test and Moran's I calculations is included as
a reference point. We see that the choropleth and bivariate map have an
I-shape, the VSUP map has an L-shape, and the pixel and transparency
maps have an upper triangular shape. The I-shape indicates that variance
has no impact on visibility, the L-shape indicates the variance only
impacts visibility at its maximum, and the upper triangular shape shows
variance consistently impacts visibility. No plot types perfectly match
the pattern in the theoretical visualization.}

\end{figure}%

Figure~\ref{fig-tileplot} shows the percent of participants who were
able to make a correct selection for each type of plot, based on the
distance (\(D\)) between the number and non-number groups, and the
standard deviation (\(V\)) of the subsamples. The grid for the
choropleth and bivariate maps appears to be quite similar, with no
visible effect from the change in \(V\), only the change in \(D\).
Looking at the VSUP map, we do get some interference, as expected, but
the interference appears to be limited to the cases when the \(D=1\) or
\(V=5\). Outside of those two scenarios for the VSUP map, there does not
appear to be a clear relationship with \(V\). The pixel and transparency
map appear to have the lower triangles of visibility. However, the
theoretical hypothesis data does not appear to have the same sensitivity
as the transparency or pixel map.

\subsection{Power analysis}\label{power-analysis}

\subsubsection{Estimating significance
levels}\label{estimating-significance-levels}

To set the \(\alpha\) for the theoretical test, we need to calculate the
\(\hat\alpha_k\) value for each plot type, \(k\). While the lineup
protocol has a theoretical calculation that can estimate \(\alpha\), as
any plot picked by chance is \(\frac 1 M\) for a lineup of \(M\) plots
(assuming no dependence structure between the plots) {[}35{]}, {[}55{]},
this calculation does not translate to uncertainty visualization. As
graphics do not come with a significance threshold {[}55{]}, and a
significance threshold is required to compare our plots to classic
hypothesis tests, we estimate \(\hat{\alpha}_k\) using the proportion of
participants who identified a number in a plot of just random noise.

\begin{figure}

\centering{

\includegraphics[width=0.9\linewidth,height=\textheight,keepaspectratio]{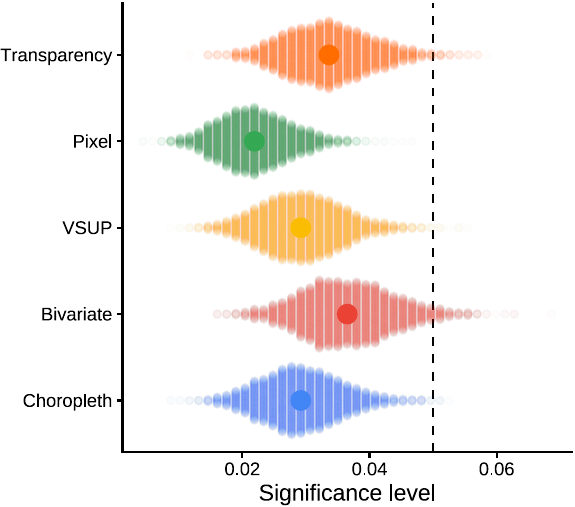}

}

\caption{\label{fig-h0calc}Bootstrapped significance levels for each
plot type, shown as jittered dotplots, computed using the null plots,
\(D=0\). The large points are the full sample values used in the power
analysis. The vertical line indicates the conventional significance
level, 0.05. All the significance levels hover around 0.03, with pixel
maps slightly lower than the others.}

\end{figure}%

Figure~\ref{fig-h0calc} shows the bootstrapped distribution for the
significance level of each plot, \(\hat{\alpha}_k\). We can see that
there is quite a large range in the distribution, which is likely caused
by individual differences in risk aversion. Reading the comments in the
study, we found that some participants thought they were always supposed
to see a number, and would make a guess if they had the slightest
inkling towards a specific number, while other participants would only
make a guess if they could fully make out the number's shape. These
tactics align with Moran's I test and \(t\)-test, respectively. It may
be more appropriate to also treat our value of \(\hat{\alpha}_k\) as a
random effects model, but this will theoretically and computationally
overcomplicate the model. Therefore, we will use the average number of
false positives within each plot type to estimate \(\hat{\alpha}_k\).

\subsubsection{Random effects model}\label{random-effects-model}

\begin{figure}

\centering{

\includegraphics[width=0.9\linewidth,height=\textheight,keepaspectratio]{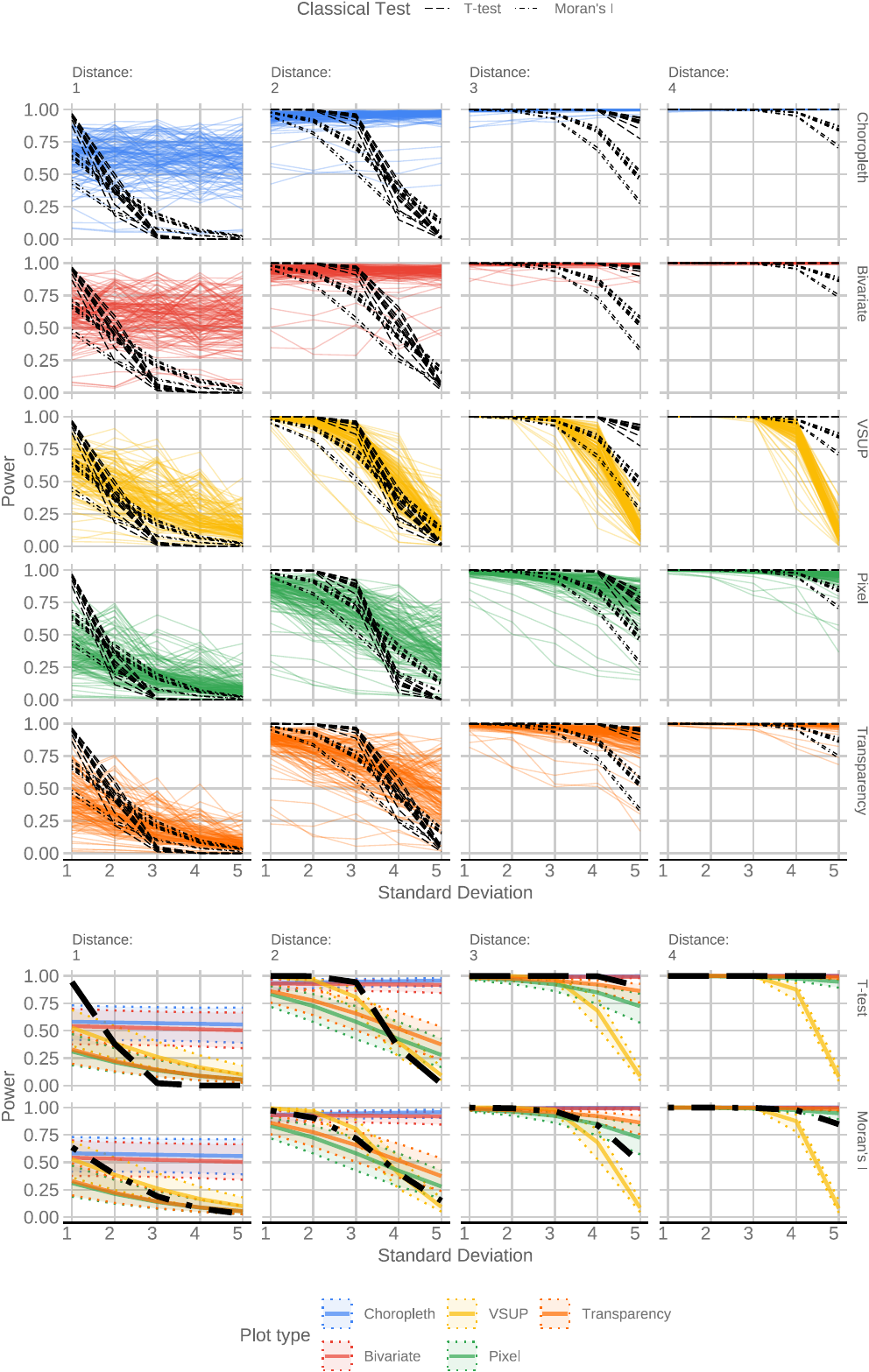}

}

\caption{\label{fig-mixmodel}The experimental power curves estimated
using a generalised linear random effects model (top) and generalised
linear fixed effects model (bottom) for all 125 experimental factors: V
(standard deviation), D (Distance), and K (plot type). Two different
dashed lines represent the results of the theoretical hypothesis tests,
the t-test, and Moran's I test. We see that the V has little impact on
the signal visibility in the choropleth and bivariate map, the signal
visibility in the VSUP map shrinks to zero when V=5, regardless of the
effect size, and the pixel and transparency map broadly follow the shape
of the theoretical tests.}

\end{figure}%

Figure~\ref{fig-mixmodel} shows the random effects model (top) and fixed
effects (bottom) components for each plot type, \(D\) and \(V\), with a
black dashed line indicating the theoretical power curve of our
hypothesis tests. The two different dashed lines show the difference
between the simulated \(t\)-test and Moran's I, and are slightly
different shapes as the power curve is set based on that plot type's
significance level, \(\hat{\alpha}_k\). The random effects model allows
us to see the variation in individual participants, while the fixed
effects average allows us to easily compare the different plot types. In
the random effects plot, the variability in the black dashed lines comes
from the random effect of the numbers shown in the plate, while the
variability in the experimental results comes from both the random
effects of the numbers in the plate and the participants' ability to
read the plot. As the variance increases, we would expect the signal in
the plate to become harder to see, as its statistical significance drops
off. Ideally, this drop off would occur at a rate that is similar to the
hypothesis tests, represented by the black dashed lines. If the coloured
line for a plot sits above the hypothesis test, then the plot is showing
a pattern that the classical hypothesis tests would consider to be
insignificant. While this can be a desirable property when comparing
standard hypothesis tests, this is not necessarily a desirable property
for uncertainty visualisation. It is important to remember that the key
issue with choropleth maps, and the reason we need uncertainty
visualisation at all, is that they \emph{always} show statistical
signal, even if that signal is spurious. We have designed this
experiment such that the visibility \emph{does} link to a signal, but
that will not always be the case. Therefore, the power curve of the most
appropriate plot will not necessarily sit above the black dashed lines,
but rather, should have a shape that is similar to the hypothesis test
curves across the different values of \(D\).

These charts largely agree with our initial findings. The choropleth and
bivariate maps are constant, and the plot visibility appears to be
largely independent of \(V\), only changing with an increase in the
average effect. The VSUP appears to follow the theoretical line for
\(D=1,2\), but diverges for \(D=3,4\) as the statistically significant
difference at \(V=5\) is hidden by the monochrome plate. This divergence
suggests that the alignment when \(D=1,2\) is a coincidence, stemming
from the fact that the theoretical test at \(D=1,2\) and \(V=5\) just
happens to produce a power of zero. Finally, the pixel and transparency
maps appear to do a reasonable job of following the power curve of the
theoretical hypothesis test, but the power is falling short at both ends
of the spectrum. The signal is not visible enough when variance is low,
but it is too visible when variance is high. This indicates that the
test is not quite sensitive enough to changing values of \(V\). This
issue in the pixel and transparency maps becomes less prevalent as \(D\)
increases, and at higher values of \(D\), the visualisation sits between
the results of the \(t\)-test and Moran's I test.

A closer inspection of the random effects models reveals some patterns
in the participants' ability to read the plot. There is variation among
participants, especially when \(D=1\). As the \(D\) increases, variation
in the participants' responses decreases, a pattern less pronounced in
the pixel and transparency map. This indicates that, as the plots become
easier to read, either due to decreasing \(V\) or increasing \(D\), the
variance among viewers also decreases. Some participants perform rather
badly across the board, but others manage to track the theoretical power
curve rather well. This indicates that, unlike in traditional hypothesis
tests, a degree of skill is involved in identifying a signal in
uncertainty visualisations. The importance of this skill becomes more
pronounced as the effect size gets smaller, indicating that training
could potentially improve the power of plots as statistical tests.

\subsubsection{Hypothesis tests}\label{hypothesis-tests}

\emph{Relating to H1: Changes to the standard deviation in the
distributions will result in no meaningful difference in our ability to
read the number in the choropleth and bivariate map.}

Table~\ref{tbl-v-trend} shows the estimated marginal effect of \(V\) for
each plot type, averaging over the effect of \(D\). If the marginal
effect is significantly different from zero, then \(V\) has some effect
on the visibility in the signal in the plot, but if it isn't
significantly different from zero, then the variance does not have any
impact on signal visibility at all. We can see that the effect
associated with \(V\) for each plot type is insignificant in the case of
the bivariate map and choropleth map, but significant for all other map
types. These conclusions remain true if we perform significance tests at
set \(D=1, 2, 3, 4\), instead of looking at the average across \(D\).
These distance-based results are available in the Supplementary
Material. This indicates that \(V\) has no significant effect on the
visibility of the signal in the bivariate and choropleth maps, given the
generalised linear mixed effects model.

\emph{Relating to H1 and H3: The probability of correctly reading the
transparency and pixel map, as well as the probability of correctly
reading the choropleth and bivariate map, will be similar.}

Comparisons of the standard deviation effect between plot types of
interest are shown in Table~\ref{tbl-basicmodel}. The \(p\)-values were
adjusted for multiple comparisons using the Sidak adjustment. If the
statistical information conveyed by two plots is equivalent, the signal
visibility between the two plots will be equivalent. If this is true,
the probability of correctly reading the transparency and pixel maps
will be similar. For the pairwise comparisons of the transparency/pixel
map and the choropleth/bivariate map, significance tests at
\(D=1, 2, 3, 4\) result in the same conclusions in terms of
significance. Distance-based results as well as all pairwise comparisons
are available in the Supplementary Material. The non-significant
difference between the transparency and pixel map, as well as between
the choropleth and bivariate map, indicates that standard deviation
effects are similar between the plot types, as expected.

\emph{Relating to H2: The interference in the VSUP map will not be
proportional to statistical significance.}

Based on Figure~\ref{fig-mixmodel}, this hypothesis is partially
supported. Values are similar to proposed power for \(D = 1, 2\), but
there is a non-proportional relationship in \(D = 3, 4\). The alignment
for \(D = 1, 2\) is likely due to alignment in the scaling of the
variance for these particular distances, but this does not hold for all
distance values.

\begin{table}

\caption{\label{tbl-v-trend}Effect of Standard Deviation (V) by Plot
Type, Averaged Over Distance (D)}

\centering{

\begin{tabular}{l|r|r|r|r}
\hline
Plot Type & V Effect & SE & Z Ratio & P-Value\\
\hline
Choropleth & 0.179 & 0.144 & 1.238 & 0.216\\
\hline
Bivariate & -0.044 & 0.121 & -0.362 & 0.718\\
\hline
VSUP & -2.486 & 0.106 & -23.359 & 0.000\\
\hline
Pixel & -0.700 & 0.058 & -12.015 & 0.000\\
\hline
Transparency & -0.604 & 0.068 & -8.843 & 0.000\\
\hline
\end{tabular}

}

\end{table}%

\begin{table}

\caption{\label{tbl-basicmodel}Selected Comparison of Plot Types for
Standard Deviation (\(V\)), Averaged Over Distance}

\centering{

\begin{tabular}{l|r|r|r|r}
\hline
Contrast & Estimate & SE & Z Ratio & P-Value\\
\hline
Choropleth - Bivariate & 0.223 & 0.188 & 1.181 & 0.475\\
\hline
Transparency - Pixel & 0.096 & 0.090 & 1.075 & 0.565\\
\hline
\end{tabular}

}

\end{table}%

\section{Discussion}\label{discussion}

All hypotheses posed at the beginning of the experiment were either
fully or partially confirmed, creating a foundation for a theory of
signal suppression in uncertainty visualisation. Unsurprisingly, the
choropleth map had an insignificant relationship with standard
deviation, as the standard deviation of each distribution was not
included in the choropleth map. Notably, the bivariate map \emph{also}
had a coefficient that was not significantly different from zero, and
its coefficients on the mixed effects model were not significantly
different from the choropleth map. This finding is of particular
interest, as bivariate maps are frequently suggested as an alternative
to choropleth maps, specifically for the purposes of suppressing
statistically invalid signals. However, these results indicate that, if
your goal is to make statistically invalid signals invisible, \emph{a
bivariate map is functionally identical to a choropleth map}. Any
benefit in using a bivariate map will need to come from explicit
calculation, and a visualisation that requires a mental calculation to
be understood defeats the purpose of making a visualisation.

The results partially supported our second hypothesis. The VSUP map,
unlike the bivariate map, did achieve visual interference and was not
functionally identical to the choropleth map; however, this interference
did not align with the theoretical hypothesis tests. The fact that every
plate at \(V=5\) was completely monochrome disproportionately pulled
down the slope of the random effects model. This caused better alignment
with the theoretical tests for \(D=2\), but incredibly poor alignment
when \(D=4\) and \(D=5\). The VSUP and bivariate maps were designed with
the implicit knowledge of the maximum and minimum values of the
data-generating process, which was only possible due to the artificial
scenario of the experiment. This requirement of knowledge could be
considered a form of data snooping that disproportionately benefits the
evaluation of the VSUP and bivariate maps, but the maps are impossible
to calculate without it, and it serves as one of the main limitations to
the approach. Usually, the scaling of our axis or colours is done
automatically, but locally scaled bivariate and VSUP maps, with no
attention to the relationship between standard deviation and estimate
value, are completely meaningless. One could argue that a workaround for
this issue is leveraging the monotonic shrinkage algorithms for VSUP
palettes suggested by Kay {[}24{]}. However, as Wickham {[}58{]} pointed
out, the distinction between a coordinate transformation and a
statistical transformation is not always clear-cut. Therefore, adjusting
the coordinate to get a specific statistical output is indistinguishable
from adjusting the statistic itself. Not only would this violate one of
our original limitations for plot selection, but it would also
significantly increase the number of plots we would need to evaluate.
Therefore, we opted to test the classic VSUP map and did not extend to
the methods suggested by Kay {[}24{]}. The VSUP approach also has
several other implicit limitations that we did not cover in this
experiment. For example, it cannot handle discontinuous distributions
that oscillate between two distant outcomes, rather than maintaining a
smooth transition between values. The VSUP approach also only works for
ordered variables, and does not make intuitive sense if we are uncertain
about a categorical fill. These limitations are absent in the pixel and
transparency map.

Finally, the results for the pixel and transparency maps have some
interesting implications for the statistical properties of plots. Given
that the two plots were aligned, we can isolate the benefit in the
approach to the display of a full distribution, rather than the
representation of a distribution as multiple values. This means we can
still effectively perform signal suppression in a situation where the
pixel map is inappropriate. These plots showed that we \emph{can}
achieve a signal that aligns with hypothesis testing (or at least a
signal that has a consistently similar shape). There are likely several
design changes that could improve this alignment, such as an optimised
colour palette or sample size.

Our results also have implications for the broader concepts of design in
uncertainty visualisations. The confirmation of \textbf{H1} indicates
that treating uncertainty as a second variable, independent of the
estimate, will not facilitate the suppression of false signals. Plotting
the estimate and uncertainty to separate aesthetic channels will prevent
the uncertainty from visually interfering with the estimate, meaning it
has no mechanism to make more uncertain estimates harder to see, and
will not prevent the visualisation from displaying spurious signals. The
confirmation of \textbf{H2} indicates that the correct representation of
an estimate should be one that completely describes the distribution.
These findings provide empirical evidence for the formalisation
suggested by Kay {[}25{]}. The confirmation of \textbf{H3} suggests
that, if we visualise our distribution as a sample, and ensure all the
draws are equally weighted, they will provide an identically valid
statistical signal. This is particularly useful for the
\texttt{ggdibbler} {[}38{]} R package, which leverages this
perpendicularity in the grammar of graphics to make a range of flexible
uncertainty visualisations for EDA. If this hypothesis had proven to be
false, these position adjustments would not be independent of the
statistical information, and changing the position adjustment would
change the signal suppression in the plot, making the suggested methods
ineffective. In the case where multiple random variables are fed into
the system at once, transparency is the only viable position adjustment
to allow for even weighting of all sample outcomes. Our results show
that the system maintains statistical validity, even when we are unable
to translate the results back to a numerical scale using the legend.

\section{Contributions, limitations, and future
research}\label{contributions-limitations-and-future-research}

Several contributions were made in this paper. We designed a
visualisation experiment that is able to measure uncertainty as a latent
variable, capturing its effect as noise, rather than signal. We also
translated the statistical approaches from the lineup protocol to
uncertainty visualisation, allowing us to compare pattern visibility to
the outcomes of a \(t\)-test and Moran's I test. By constructing the
plots with respect to the grammar of graphics, we were able to
understand \emph{why} some approaches may, or may not, work and
contribute towards a general theory of uncertainty visualisation.

This work is not without its limitations, the most substantial being the
restriction of visual encoding to colour. Re-sampling approaches are
flexible, and capable of visualising uncertainty from multiple channels
simultaneously {[}37{]}, {[}38{]}, {[}42{]} which is a much better
reflection of how uncertainty typically appears in analysis. Contrary to
this, our evaluation method is restricted to colour and colour alone. In
this case, we were able to leverage existing colour blind tests to
evaluate uncertainty as a latent variable, but every visual aesthetic
does not have such a convenient equivalent. Aesthetics such as shape or
text could have a similar number reading test (where the shape is the
number), but position, size, or length do not have an obvious
equivalent. Developing approaches that work for different or multiple
aesthetics would require substantial creativity, and/or a more
fundamental understanding of what it means to extract information from
specific visual channels.

Another limitation is that we were strict in our data-generating
process. We kept the data generation very simple to avoid confounding
data issues with other untested aspects of the design, but this resulted
in strict assumptions. Our distribution shapes were limited to normal
distributions, every observation had an identical standard deviation
with no interfering pattern, and the only variance between observations
came from the estimate's central values. While this data-generating
process is limited, deviations from this strict scenario are more likely
to help, rather than hinder, the uncertainty visualisation in comparison
to the theoretical hypothesis tests. There is ample evidence that
classical hypothesis tests will outperform visual inference when the
assumptions of the test are true, but the visual tests will do better
when the assumptions are false {[}20{]}, {[}30{]}, {[}35{]}. This
indicates that more unusual data-generating processes that are more
likely to violate traditional test assumptions may improve the
performance of the visualisations.

There were also some plot design factors that likely impact the
readability of the figures, but were outside the scope of this
experiment. Anecdotally, when designing the experiment, we found the
palette had an effect on the signal visibility. A different choice in
colour palette may produce different results, although we hypothesise
that the standard deviation effects would remain similar. Therefore,
values resulting from visibility calculations should not be treated as
definitive values for the data used to generate the plots.

There are several extensions to this work that might improve the
sensitivity issues in the pixel and transparency maps, such as the
effect of sample size. We set the number of samples within each plot to
50, which reduces the variance within the visualisation, but also
compresses the colour scale, making individual colours harder to
discern. There may be a sample size trade-off in designing these
visualisations effecting their sensitivity. Theoretically, the pixel map
approach could be generalised to any visualisation that can be made in
\texttt{ggplot2} {[}38{]}, so the signal suppression approach could be
investigated for other aesthetics in other plot types. Additionally, we
could extend our question to cases where we have multiple uncertain
variables at once, rather than only allowing variability in colour. The
plot size also seemed to have an effect on the visibility of the
pattern, with smaller plots making the pattern easier to see. Since the
transparency map was effective, it would become a viable visualisation
approach if an interpretable legend could be presented alongside the
map. In its current state, as can be seen in Figure~\ref{fig-maps}, as
the standard error increases, the colours quickly diverge from those in
the legend.

There are also several potential extensions or issues we encountered
when implementing the colour blind test that should be of note to anyone
who wants to implement a similar method. We noticed an afterimage
occasionally when doing the test, so including an image as a mask
between plots may help alleviate this issue. We did not conduct testing
of font types, so the tendency for participants to get confused between
numbers (specifically 6, 3 and 8) might be mitigated with a better font
choice.

Finally, while the Ishihara test was chosen due to its similarity to
spatial pattern identification, the tasks are not identical. Alternative
colour blind tests that ask participants to trace a shape, rather than
identify a number, may reduce memorisation and better reflect the
objectives of choropleth maps. This could be achieved with software,
such as the \texttt{r2d3} {[}45{]} R package, which allows participants
to draw on data visualisations.

\section*{Ethics declaration}\label{ethics-declaration}
\addcontentsline{toc}{section}{Ethics declaration}

Ethics approval for the online survey was granted by Monash University
Human Research Ethics Committee (Project ID 51214). All applicants
provided informed consent prior to participating in this research.

\section{Acknowledgements}\label{acknowledgements}

The first author of this paper is supported in part by a scholarship
from the Australian Energy Market Operator. This research was supported
by the Commonwealth through an Australian Government Research Training
Program Scholarship {[}DOI: https://doi.org/10.82133/C42F-K220{]}. We
thank Susan VanderPlas, Sarah Goodwin, Emily Robinson, and our anonymous
reviewers for their insightful comments and feedback, which
substantially improved the work. We also thank Janith Wanniarachchi for
his help troubleshooting shiny server issues. The R packages used for
this work were: \texttt{tidyverse}, \texttt{distributional},
\texttt{ggdist}, \texttt{ggdibbler}, \texttt{patchwork},
\texttt{khroma}, \texttt{colourspace}, \texttt{ozmaps}, \texttt{sf},
\texttt{ggthemes}, \texttt{MASS}, \texttt{shadowtext},
\texttt{flextable}, \texttt{emmeans}, \texttt{kableExtra},
\texttt{broom}, \texttt{lme4}, \texttt{car}, \texttt{janitor},
\texttt{packcircles}, \texttt{gglogo}, \texttt{scales}, \texttt{glue},
\texttt{digest}, \texttt{ggbeeswarm}, \texttt{conflicted}, \texttt{sp},
and \texttt{spdep}. The GitHub repository for this paper can be found at
https://github.com/harriet-mason/uncertainty-experiment, which contains
the files required to reproduce this article in full.

\section{References}\label{references}

\protect\phantomsection\label{refs}
{\fontsize{8pt}{9.6pt}\selectfont

}

\end{document}